\documentclass[prb,twocolumn,showpacs,preprintnumbers,amsmath,amssymb]{revtex4-1}
\usepackage{amssymb}
\usepackage{bm}
\usepackage{amsmath}
\usepackage[dvips]{graphicx}
\usepackage{color}
\usepackage{comment}
\usepackage{multirow}
\usepackage{subfigure}

\newcommand \cblack {\color{black}}

\begin{document}
\color{red}
\title{Semiclassical Landau quantization of spin-orbit coupled systems}
\color{black}

\author{Tommy Li$^1$, Baruch Horovitz$^2$, Oleg P. Sushkov$^1$}
\affiliation{$^1$School of Physics, University of New South Wales, Sydney 2052, Australia}
\affiliation{$^2$Department of Physics, Ben Gurion University, Beer-Sheva 84105, Israel}

\begin{abstract}
 A semiclassical quantization condition is derived for Landau levels in general spin-orbit coupled systems. This generalizes the Onsager quantization condition via a matrix-valued phase which describes spin dynamics along the classical cyclotron trajectory. We discuss measurement of the matrix phase via magnetic oscillations and electron spin resonance, which may be used to probe the spin structure of the precessing wavefunction. We compare the resulting semiclassical spectrum with exact results which are obtained for a variety of spin-orbit interactions in 2D systems.
\end{abstract}
\maketitle

\section{Introduction}

Two-dimensional (2D) semiconductor systems offer strong and tunable intrinsic spin-orbit interactions\cite{Engels,Grbic} which have been exploited in recently proposed spintronic devices \cite{DattaDas,Schliemann,Fabian,ZuticSpintronics}. In these systems, the close relationship between charge and spin dynamics produces strongly modified electronic transport properties which are exhibited in a range of effects including the anomalous Hall effect\cite{NozieresAHE,JungwirthAHE,NagaosaRevAHE}, the spin Hall effect \cite{MurakamiSHE,SinovaSHE,BernevigSHE,SchliemannSHE} and weak antilocalization \cite{BergmannWAL}. In the past, quantum interference measurements have been proposed as a probe of the spin-orbit interaction in low-dimensional semiconductor heterostructures due to their sensitivity to quantum phases arising from coherent spin precession accompanying ballistic transport \cite{AronovLyandaGeller,ArovasLyandaGeller,YauPhase,Nagasawa,LiSushkov}. In particular, the role of adiabatic and non-adiabatic phases in magnetic oscillations \cite{WinklerShayegan,KeppelerWinkler} is of high interest due to the fact that oscillatory magnetotransport experiments have provided crucial measurements of the spin-orbit coupling in these systems\cite{Eisenstein,Engels,LuoSDH1,DorozhkinSDH,LuoSDH2,SchultzSDH,RamvallSDH,HeidaSDH,HuSDH,GrundlerSDH,MatsuyamaSDH,Grbic}. Furthermore, recent experimental and theoretical studies of 2D Dirac systems such as graphene and surface states of three-dimensional topological insulators have highlighted the role of the geometric phase in particular in magnetotransport \cite{Novoselov,Zhang,Analytis,Xiu,Xiong,Sacepe,Fuchs,Wright1,Wright2,Goerbig}.

In this work we derive an expression for the Landau level spectrum of a 2D system with spin-orbit interaction via a generalization of the Onsager quantization condition\cite{Onsager} to account for non-trivial spin dynamics. Spin evolution is encoded in the SU(2) phase representing the total rotation of an initial spin state around a period of cyclotron motion. This SU(2) phase is necessary to describe non-adiabatic spin dynamics which is present when the effective magnetic field in momentum space describing the spin-orbit interaction is not sufficiently strong to locally polarize the spin of the particle along the orbit\cite{LiSushkov,LiYeoh}. We evaluate the semiclassical spectrum for the cases when the spin-orbit effective magnetic field is simply rotating in momentum space with  a single winding number and compare to the exact solutions for a variety of spin-orbit interactions in semiconductor systems, including several cases which have not been previously mentioned in the literature.
In addition, we show that magnetic oscillations and electron spin resonance (ESR) serve as effective probes of the precessing spin structure of Landau level states.

This paper is organized as follows: in Section II we derive the semiclassical quantization condition and a general expression for the Landau level spectrum of spin-orbit coupled system, accounting for spin dynamics via a matrix valued phase. In Section III we discuss magnetic oscillations and derive the expression for the oscillatory density of states in terms of the matrix-valued phase.
In Section IV we evaluate the level spectrum and eigenstates for a rotating spin-orbit interaction with fixed winding number. We also calculate exact results for a variety of interactions in $p$-type systems and present a comparison of the semiclassical and exact results for these as well as for previously discussed results in $n$-type systems.\cite{ZareaUlloa,ZhangLandau,PappMicu,Erlingsson}
In Section V we discuss ESR as a probe of the spin-orbit interaction type and evaluate the ESR matrix elements for the cases discussed in Section IV. Our summary and concluding remarks are presented in Section VI.

\section{Spectrum of Landau levels with spin-orbit interaction}

We consider a 2D electron or hole gas in perpendicular magnetic field, described by the Hamiltonian
\begin{gather}
H = \frac{ \bm{\pi}_x^2 + \bm{\pi}_y^2}{2m} + H_s(\bm{\pi}_x, \bm{\pi}_y) \ \ , \ \  H_s = \beta(\bm{\pi}_x, \bm{\pi}_y ) \cdot\sigma
\label{hamil}
\end{gather}
where $\bm{\pi} = (\bm{\pi}_x, \bm{\pi}_y) =\bm{p}- e \bm{A}$ are the operators of kinetic momentum\cite{note} ($e$ is the charge of the electron or hole) and $m$ is the effective mass. For electron systems, the Pauli matrices $\sigma$ act on spin, while for hole systems, $\sigma$ acts on the doublet of heavy hole states\cite{WinklerBook}. The spin-dependent interaction $H_s$, accounting for the spin-orbit interaction in addition to Zeeman coupling to the external magnetic field, will be expressed in terms of the effective magnetic field in momentum space, $\beta = (\beta_x, \beta_y, \beta_z)$.

Since the kinetic momenta satisfy a commutation relation $\left[\bm{\pi_x}, \bm{\pi_y} \right] = i e B_z$,  we may construct annihilation and creation operators $\bm{a},\bm{a}^\dagger$ with $\bm{a} = \frac{ \bm{\pi}_x + i \eta \bm{\pi}_y}{|2e B_z|}$ where $\eta$ is the sign of the charge,  with the corresponding number operator
\begin{gather}
\bm{N} = \bm{a}^\dagger \bm{a} =  \frac{ \bm{\pi}_x^2 + \bm{\pi}_y^2}{|2e B_z|} - \frac{1}{2} \ \ .
\label{number}
\end{gather}
It is possible to diagonalize the Hamiltonian (\ref{hamil}) in the number representation, as has been done in previous approaches to the problem\cite{ZareaUlloa,ZhangLandau,PappMicu,Erlingsson}. We consider however the semiclassical picture, in which the spectrum is related to the dynamics of wavepackets moving along the cyclotron trajectory. Due to the spin-orbit interaction, a wavepacket in some initial polarization state will precess along the orbit, and generally undergo a rotation after a complete revolution, which is described by an SU(2) matrix. Thus for the purposes of semiclassical quantization the phase is matrix-valued, and the spectrum will be determined by the eigenvalues and eigenvectors of the matrix-valued phase. In order to rigorously derive this result, we introduce the spinor wavefunction $\psi(\theta)$ which varies along the angle $\theta$ in momentum space. Explicitly, this is given by $\psi(\theta) = \langle \theta | \psi \rangle$ where the states $|\theta\rangle$ are related to number eigenstates via\cite{Carruthers}
\begin{gather}
|\theta \rangle = \sum_{n = 0}^{\infty}{ e^{i \eta n \theta} |n\rangle} \ \ , \ \ \eta = \text{sgn}(e) \ \ .
\label{basis}
\end{gather}
where $|n\rangle$ are eigenstates of the number operator. Here we assume $B_z > 0$. The basis states $|\theta\rangle$
are eigenstates of the operators $e^{i\bm{\theta}}, e^{-i \bm{\theta}}$ which are related to the momentum operators via
\begin{gather}
\bm{a} = e^{i \bm{\theta}} \sqrt{\bm{N}} = \frac{ \bm{\pi}_x + i \eta \bm{\pi}_y}{|2e B_z|} \ \ .
\end{gather}

Note that the number operator corresponds to the classical action coordinate, while $\theta$ 
is simply the angle in momentum space, $(\pi_x, \pi_y) = (|\pi| \cos \theta, |\pi| \sin\theta)$. Thus in the semiclassical limit the wavefunction $\psi(\theta)$ represents the motion of a particle in momentum space as a function of the angle $\theta$.

In order to obtain the Schr\"{o}dinger equation for $\psi(\theta)$ then obtained from the Hamiltonian (\ref{hamil}), we note that the classical coordinates $(\theta, N)$ are canonically conjugate, which implies that the operator ${\bm N}$ takes the form of a derivative operator in the $\theta$ representation. Explicitly acting on the basis (\ref{basis}) with $\bm{N}$ shows that
\begin{gather}
\langle \theta| \bm{ N} | \psi \rangle = i \eta \frac{d}{d\theta} \langle \theta | \psi \rangle = i \eta \frac{d\psi(\theta)}{d\theta} \ \ .
\end{gather}
Thus the first term in (\ref{hamil}) may be replaced with $ \frac{\bm{\pi_x}^2 + \bm{\pi_y}^2}{2m} \rightarrow \omega (i \eta \frac{d}{d\theta} + \frac{1}{2})$ where $\omega = |\frac{e B_z}{m}|$ is the cyclotron frequency, and the effective magnetic field may be regarded as a function of the coordinates $(\theta, \bm{N} \rightarrow i \eta \frac{d}{d\theta})$. Thus the Schr\"{o}dinger equation for $\psi(\theta)$ reads:
\begin{gather}
\left[ i \eta \omega\frac{d}{d \theta} + \beta(\theta, i \eta \frac{d}{d \theta}) \cdot \sigma  - E + \frac{\omega}{2} \right] \psi(\theta) = 0 \ \ , \ \  \omega = \frac{|e B_z|}{m} \ \ .
\label{Schrodinger}
\end{gather}
In the absence of spin-orbit interaction, $\beta = 0$, the wavefunction satisfies the equation
\begin{gather}
i \eta  \frac{d \psi}{d\theta} =  \nu \psi, \ \ \nu = \frac{E}{\omega} - \frac{1}{2}
\end{gather}
which yields the  wavefunction
\begin{gather}
\psi(\theta) = e^{- i \eta \nu \theta} \ \ ,
\end{gather}
corresponding to a circular orbit in momentum space, with $\eta = \text{sgn}(e)$ indicating the direction in which the circle is traversed (clockwise for $\eta > 0$ and anticlockwise for $\eta<0$). (This is simply the wavefunction of the harmonic oscillator in the phase representation.\cite{Carruthers}) Single valuedness of the wavefunction then requires $\nu$ to be an integer, which of course yields the usual Landau level spectrum $E_n = (n + \frac{1}{2}) \omega$. However, $\nu$ is also related to the average momentum of the orbit via
\begin{gather}
\langle \bm{\pi}^2\rangle = |2e B_z|( \langle \bm{N} \rangle + \frac{1}{2} ) 
 = |2e B_z| ( \nu + \frac{1}{2}) \ \ .
\end{gather}
Thus single valuedness of the wavefunction in the $\theta$-representation is equivalent to Onsager's quantiztion rule\cite{Onsager}, that the area of the momentum space orbit must be quantized:
\begin{gather}
\frac{1}{|2e B_z|} \int_0^{2\pi}{ \pi^2  d\theta} = 2\pi(n + \frac{1}{2}) \ \  .
\label{Onsager}
\end{gather}


In the presence of spin-orbit coupling the spinor  $\psi(\theta)$  generally precesses as a function of $\theta$ under the influence of the effective magnetic $\beta(\theta, N)$. In the semiclassical r\'{e}gime, $\langle \bm{N} \rangle \gg 1$ the wavefunction takes the form of a Born-Oppenheimer product of orbital and spin factors,
\begin{gather}
\psi(\theta) = e^{-i \eta \nu \theta} \chi(\theta) \ \ .
\label{BornOppenheimer}
\end{gather}
The first factor in (\ref{BornOppenheimer}) corresponds to an orbital trajectory in momentum space with radius $|\pi| = \sqrt{|2e B_z|(\nu + \frac{1}{2})}$ (and we assume $\chi^\dagger \chi = 1$).
When $\chi(\theta)$ is slowly varying compared to the orbital factor, we may replace the action of the derivative with the semiclassical variable $\nu$ in $\beta(\theta, i \eta \frac{d}{d\theta}) \rightarrow \beta(\theta, \nu)$. 
This requires that the spin-orbit effective magnetic field be much smaller than the total energy, $|\beta(\nu, \theta)|\ll E$, as well as $\frac{d\chi}{d\theta} \ll \nu $. Nevertheless, this does not require $|\beta(\nu, \theta)| \ll \omega$  (this inequality is violated, e.g. in the r\'{e}gime of double magnetic focusing\cite{Rokhinson}).  
Thus the Schr\"{o}dinger equation for spin reads
\begin{gather}
-i \eta \omega \frac{d \chi}{d\theta} = \left[ \beta(\theta, \nu) \cdot \sigma - \omega \delta \right]\chi
\label{spinevolution}
\end{gather}
where $\delta$ is a parameter defined by
\begin{gather}
E = \omega( \nu + \frac{1}{2} + \delta) \ \ .
\label{delta}
\end{gather}
Equation (\ref{spinevolution}) is identical to the equation of motion for a precessing wavepacket moving along the classical cyclotron orbit,  $(\nu = const., \theta = - \eta \omega t)$ if the left hand side is replaced by the time along the trajectory, $-i \eta \omega \frac{d}{d\theta} = i \frac{d}{dt}$. We may divide the evolution of spin into two parts, $\chi(\theta) = e^{-i \eta \delta \theta} U(\theta) \chi(0)$ where $U(\theta)$ is an SU(2) matrix,
\begin{gather}
U(\theta) = \mathcal{P} e^{ i \eta \omega^{-1}\int_0^\theta{ \left[ \beta(\nu, \theta) \cdot \sigma \right] d\theta}} \ \ ,
\label{U}
\end{gather}
where $\mathcal{P}$ indicates path-ordering. Over a complete orbit, the spin wavefunction accumulates a complex phase factor, as well as a rotation generated by the matrix-valued phase $U(2\pi)$; however in a stationary state, the spin polarization must return to its initial value after a complete orbit, implying that the spinors $\chi(0)$ and $\chi(2\pi)  = e^{- 2\pi i \eta \delta }U(2\pi) \chi(0)$ differ at most by a phase. It follows that $\chi(0) = \chi_\pm$ is an eigenvector of $U(2\pi)$. Since $U(2\pi)$ is an SU(2) matrix, its eigenvalues $e^{+i \Phi}, e^{-i \Phi}$ are complex conjugate, and the phase accumulated due to unitary transformations of spin over a complete orbit for an initial spin state $\chi_\pm$ is equal to $e^{\pm i \Phi}$. The wavefunction $\psi(\theta)$ must be single valued, implying the quantization condition
\begin{gather}
2\pi (\nu + \delta) \pm \Phi  = 2\pi n \ \ ,
\label{quantization}
\end{gather}
Recalling the definition (\ref{delta}), this gives a relationship between the spectrum and the eigenvalues of the matrix phase $U(2\pi)$ in the semiclassical limit,
\begin{gather}
E_{n, \pm} = \omega( n + \frac{1}{2} \pm \frac{\Phi}{2\pi}) \ \ .
\label{spectrum}
\end{gather}
The matrix phase is fully determined by the path ordered exponential (\ref{U}) which depends on the radius of the orbital trajectory, $|\pi| = \sqrt{|2e B_z| (\nu + \frac{1}{2})}$.
Nevertheless, the quantization condition (\ref{quantization}) does not fix $\nu$ and $\delta$ individually, but only the combination $\nu + \delta$, with $\delta$ being a free parameter corresponding to an arbitrary choice of the phase of $\chi$. The choice of $\delta$ is fixed by the requirement for the validity of the Born-Oppenheimer approximation, $\frac{ d\chi}{d\theta} \ll \nu $. This requires $\delta \ll \nu$. Since the total phase accumulated by $\chi$ in a stationary state is $2\pi \delta \pm \Phi$, we must minimize the variation of $\chi$ along the trajectory by choosing the phase of $\chi$ so that $\nu$ is an integer,
\begin{gather}
\delta = \mp \frac{ \Phi}{2\pi} \ \ , \ \ \nu = n \ \ .
\end{gather}
Thus the spectrum is determined by the eigenvalues of the matrix phase $U(2\pi)$ evaluated for orbits in momentum space with radius $|\pi_n| = \sqrt{|2e B_z|(n + \frac{1}{2})}$. Note that in the absence of electric fields the Hamiltonian (\ref{hamil}) commutes with the guiding center operators $\bm{X} = \bm{x} + \frac{\bm{\pi}_y}{e B_z} \ \ , \ \ \bm{Y} = \bm{y} - \frac{\bm{\pi}_x}{e B_z}$, thus each Landau eigenstate may be chosen to be a simultaneous eigenstate of $\bm{X},\bm{Y}$. This leads to the usual degeneracy per unit area  $\frac{|e B_z|}{2\pi}$.

The quantization condition (\ref{quantization}) may be expressed in terms of the energy,
\begin{gather}
\mathcal{J}_\pm (E) = \frac{2\pi m E}{|e B_z|} \mp \Phi = 2\pi(n_\pm(E) + \frac{1}{2}) \ \ .
\label{action}
\end{gather}
The left-hand side of (\ref{action}) is equal to total phase accumulated by the wavefunction $\psi(\theta)$ in a stationary state, and is therefore equal the classical action integrated over a single period. At a given energy, there exist two orbits, whose radii in momentum space are given by (from (\ref{delta}))
\begin{gather}
\frac{1}{|2e B_z|} \int{ \pi_\pm^2 d\theta} = 2\pi (n_\pm(E) + \frac{1}{2}) 
\end{gather}
and may be determined \emph{e.g.} from magnetic focusing.\cite{Rokhinson} The periods of the two spin trajectories are given by the derivative of the action with respect to the energy,
\begin{gather}
T_\pm =
\frac{2\pi}{\omega_\pm} = \frac{d\mathcal{J}_\pm}{dE} = \frac{ 2\pi m}{|e B_z|} \mp \frac{ d\Phi}{dE} \ \ ,
\label{omega}
\end{gather}
regarding $E$ as a continuous variable in the semiclassical limit.

While we have performed a detailed derivation in the case of a quadratic dispersion, it is intuitively clear that our argument and results may be rigorously generalized to the case of non-quadratic dispersions, $\frac{ \pi^2}{2m} \rightarrow \epsilon(\pi)$. In this case the wavefunction (\ref{BornOppenheimer}) takes the form $\psi(\theta) = e^{- \frac{i \eta }{|2e B_z|} \int{ \pi^2 d\theta} + \frac{ i \eta\theta}{2}} \chi(\theta)$ where $\chi(\theta)$ satisfies the same Schr\"{o}dinger equation, (\ref{spinevolution}) with the spin-orbit interaction $\beta(\pi_x, \pi_y)$ evaluated for quantized orbits of constant energy satisfying the condition $\frac{1}{|2e B_z|}\int{ \pi_n^2 d\theta} = 2\pi(n + \frac{1}{2})$.
The spectrum (\ref{spectrum}) for non-quadratic dispersions becomes
\begin{gather}
E_{n, \pm} = \epsilon(\pi_n) + \omega( \frac{1}{2} \pm \frac{\Phi}{2\pi}) \  \ ,
\end{gather}
where the oscillator frequency $\omega$  must be calculated from the classical equations of motion corresponding to the general dispersion $\epsilon(\pi)$.
\cblack

\section{Magnetic Oscillations}

According to Onsager's principle, the oscillations in resistivity of a 2D system as function of perpendicular magnetic field directly measure the semiclassical phase (\ref{action}) accumulated over an orbit for a particle at the Fermi energy.\cite{Onsager}
We will calculate the oscillating resistivity for a general spin-orbit coupled system  in  a similar manner to the method of Lifshitz and Kosevich \cite{LifshitzKosevich}.
In the Drude approximation the conductivity is proportional to the density of states
\begin{gather}
A(E) = - \frac{|e B_z|}{4\pi^2} \text{Tr} G^R = - \frac{|e B_z|}{4\pi^2} \text{Im} \sum_{n, \sigma = \pm}{\frac{1}{ E - E_{n \sigma} + \frac{i}{2\tau} } }
\label{A}
\end{gather}
where the retarded Greens function $G^R$ is averaged over disorder, and we have included the Landau level degeneracy factor $\frac{|e B_z|}{2\pi}$. We only consider the situation where impurities are short-ranged, so that relaxation is described in first order by a single parameter $\tau^{-1}$ equal to the total scattering cross section at\cite{tau} $B_z = 0$.
Applying the Poisson summation formula to (\ref{A}), one regards $E_{n \pm} \rightarrow E_\pm(\mathcal{J})$ as a function of the continuous variable $\mathcal{J}$ (\ref{action}): 
\begin{gather}
A(E) = - \frac{|e B_z|}{4\pi^2} \text{Im} \sum_{l = 0, \sigma = \pm}^{\infty} { \int{ \frac{e^{-i l (\mathcal{J}- \pi)}}{E - E_\sigma(\mathcal{J}) + \frac{i}{2\tau}} d\mathcal{J} } } \ \ .
\end{gather}
Performing a change of variables and a contour integration gives the density of states at $E = E_F$
\begin{gather}
A(E_F) = 
 \frac{|e B_z|}{2\pi} \sum_{l = 0, \sigma = \pm}^{\infty}{ \frac{1}{\omega_\sigma} e^{- \frac{ \pi l}{ \omega_\sigma \tau}} \cos\left[l \mathcal{J}_\sigma(E_F) - l\pi \right]}  \ \ ,
 \nonumber \\
  =\frac{|e B_z|}{2\pi} \sum_{l = 0, \sigma = \pm}^{\infty}{ \frac{1}{\omega_\sigma} e^{- \frac{ \pi l}{ \omega_\sigma \tau}} \cos l \left[\frac{2 \pi E_F}{\omega} - \pi - \sigma \Phi(E_F) \right]}  \ \ ,  
\label{osc1}
\end{gather}
where $\omega_\sigma = \omega_\pm$ are the frequencies of the spin trajectories (\ref{omega}). \cblack
The spin-dependent phase shift in magnetic oscillations is therefore equal to $+\Phi, -\Phi$ for orbits evaluated at the Fermi energy.
 In the semiclassical regime, $n \gg 1$, the difference between $\omega_+, \omega_-$ may be neglected in the first approximation (typically oscillations are observed up to $n \approx 40$ in electron systems \cite{Engels} and $n \approx 20$ in hole systems \cite{Grbic,LiYeoh}). Accounting for only the first harmonic in (\ref{osc1}), the resistivity becomes
\begin{gather}
\rho_{xx}(B_z) = \rho_{xx}(0)(1 + e^{ -\frac{\pi}{\omega \tau}} \cos \Phi(E_F) \cos 2\pi \left[ \frac{E_F}{\omega} - \frac{1}{2} \right]) \ \ ,
\label{env}
\end{gather}
and the oscillatory part vanishes when $\Phi(E_F) = \pi (n + \frac{1}{2})$. Since the spin-orbit interaction is generally highly tunable by experimental parameters\cite{Engels,Grbic}, measurement of the envelope $\cos \Phi$ over a range of parameters would permit the indirect mapping of semiclassical spin dynamics along the cyclotron trajectory (as we shall demonstrate in Section IVC).

\subsection{Berry phase}

In the typical experimental situation reported in magnetotransport measurements in $n$-type narrow gap systems,\cite{Engels,Eisenstein,LuoSDH1,LuoSDH2,SchultzSDH,RamvallSDH,HeidaSDH,HuSDH,GrundlerSDH} the spin-orbit interaction is sufficiently strong $|\beta(\nu, \theta)| \gg \omega$, that spin precession is adiabatic, i.e. the spin polarization is locally aligned with the vector $\beta(\nu, \theta)$ along the cyclotron orbit. In this r\'{e}gime the phase $\Phi$ contains a Berry phase\cite{Berry} contribution $\varphi_B$ equal to $-\frac{1}{2} \times$ the solid angle enclosed by the precessing spin polarization on the sphere. While this contribution has been experimentally observed\cite{Zhang,Analytis,Xiu,Xiong} and theoretically studied\cite{Fuchs,Wright1,Wright2,Goerbig} in the context of 2D Dirac materials, measurement of the Berry phase via magnetotransport in semiconductor systems has yet to be reported. Nevertheless, Eq. (\ref{env}) demonstrates that the Berry phase should appear as a correction to the phase of the resistivity oscillations, typically of order $\pi$ for strong spin-orbit interaction, which may significantly alter the amplitude of the oscillating resistivity. In the case of a strong Rashba interaction, the Berry phase $\varphi_B = \pi$ is a constant shift corresponding to a phase inversion of the oscillations. The phase may be expressed in terms of the spin-split densities measured at zero magnetic field,
\begin{gather}
\mathcal{J}_\pm(E_F) \rightarrow \frac{ 4 \pi \rho_\pm}{|2e B_z|} \mp \varphi_B \ \ .
\label{adiabaticphase}
\end{gather}
 When the Berry phase is constant as a function of the perpendicular field, it does not affect the spin-split densities $\rho_\pm$ which are usually extracted by performing a Fourier transform of the resistivity. In this case, the Berry phase appears only as a constant shift of the oscillations. Explicitly, the Fourier transform of the resistivity with respect to the inverse magnetic field is given by
\begin{gather}
\mathcal{F}(r) = \int{ e^{i b r} \rho_{xx}(b) db} \ \ , \ \ b = \frac{8 \pi^2}{|2e B_z|}
\end{gather}
and the maxima of the function $\mathcal{F}(r)$ occur at
\begin{gather}
r = r_+, r_- \ \ ,  \ \ r_\pm = \frac{\rho}{2} \mp \frac{ \partial \Phi}{\partial b} = \rho_\pm \pm \frac{ \partial \varphi_B}{\partial b} \ \ .
\label{maximafourier}
\end{gather}
Note that we assume that the spin-orbit interaction is held constant while $B_z$ is varied. While in general, the derivative of the Berry phase appears as a correction to the peaks of the Fourier transform (alongside the zero-field densities $\rho_\pm$), in the limit of strong spin-orbit interaction typically encountered in narrow-gap semiconductors\cite{Engels,Eisenstein,LuoSDH1,LuoSDH2,SchultzSDH,RamvallSDH,HeidaSDH,HuSDH,GrundlerSDH}, the Berry phase is a constant shift and does not  contribute to the position of the peaks: a Fourier analysis of the oscillations gives only the zero-field densities $\rho_\pm$, which corresponds to the first term in (\ref{adiabaticphase}). Comparison to the envelope (\ref{env}) can therefore directly reveal the Berry phase shift.

In the general situation, determination of the zero-field densities $\rho_\pm$ from a Fourier analysis of the oscillations is not straighforward due to the presence of the derivative of the Berry phase in (\ref{maximafourier}). In this case, measuring the oscillations at sufficiently low fields for which only one species contributes to the resistivity\cite{Eisenstein} would allow the Berry phase to be simply extracted from the positions of the maxima of the oscillations (as it is, \emph{e.g.} in Dirac semimetals\cite{Zhang,Analytis,Xiu,Xiong}).

\section{The case of rotating interactions: comparison of exact and semiclassical solutions}

In the typical experimental situation, semiconductor heterostructures are subject to the Rashba\cite{Rashba} and Dresselhaus\cite{Dresselhaus} spin-orbit interactions, in addition to applied magnetic fields. The competition between these interactions, which are often of the same order\cite{Ganichev,Averkiev,Meier} result in complex spin trajectories which are reflected in both the spectrum and magnetic oscillations via the spin evolution matrix $U(2\pi)$ (\ref{U}). 
Nevertheless, an important situation arises when a single spin-orbit interaction is present which corresponds to a field $\beta$ which performs an integer number of rotations $W$ around a circle in momentum space,  $\beta = (\beta_\parallel \cos (W \theta + \phi), \beta_\parallel \sin (W \theta + \phi), \beta_z)$.  We consider the realisation of this situation in both electron and hole systems. In electron systems, a pure Rashba interaction, $H_R = \alpha (\bm{\pi}_x \sigma_y - \bm{\pi}_y \sigma_x)$ corresponds to winding number $W = +1$, a pure Dresselhaus interaction in zincblende systems confined perpendicular to a cubic axis, $H_D = \alpha(\bm{\pi}_x \sigma_x - \bm{\pi}_y \sigma_y)$ corresponds to winding number $W = -1$. 

In hole systems, the higher angular momentum $J = \frac{3}{2}$ for holes implies that interactions linear in $J_x, J_y$ have higher winding numbers than their counterparts in electron systems. The Rashba and Dresselhaus interactions correspond to $W = +3$ and $W = +1$ respectively, and an applied in-plane magnetic field corresponds to $W = +2$. This statement may be derived from the observation that, for hole systems confined to a two-dimensional plane, the low energy subspace consists of the heavy hole doublet with angular momentum quantized along the perpendicular axis, $|+\rangle = |J_z = \frac{3}{2} \rangle, |-\rangle = |J_z = -\frac{3}{2}\rangle$. Interactions which are linear in $J_x, J_y$ do not couple states $|+\rangle, |-\rangle$. In order to obtain a coupling between these states, it is necessary to account for the additional interaction $\propto \bm{\pi}_+^2 J_-^2 + h.c.$ which appears in the Luttinger Hamiltonian\cite{Luttinger}. In combination with interactions linear in $J_x, J_y$, this contributes a factor $\bm{\pi}_+^2$ which, after projection onto heavy hole states raises the winding number by 2. In perturbation theory one obtains, for an in-plane magnetic field a Hamiltonian $\propto B_+ \bm{\pi}_+^2 \sigma_- + h.c.$ with $W = +2$; for the Rashba interaction the Hamiltonian is $\propto i\bm{\pi}_+^3 \sigma_-$ with $W = +3$, and for the Dresselhaus interaction the Hamiltonian is $\propto \{\bm{\pi}^2_+, \bm{\pi}_- \}\sigma_- + h.c.$ with $W = +1$.
In this section we present analytical results for these situations, and compare the semiclassical approximation to the exact spectra obtained from brute force diagonalization.

Let us first consider the situation for a general winding number $W$. The interaction with the effective magnetic field may be written $\beta\cdot \sigma = g (\beta_0 \cdot \sigma) g^{-1}$ with $g = e^{- \frac{ i W \sigma_z \theta}{2} }$ and  $\beta_0 = (\beta_\parallel \cos \phi, \beta_\parallel \sin \phi, \beta_z)$  constant along the circular trajectory. Performing a transformation to the corotating frame, $ \chi = g \chi'$, the Schr\"{o}dinger equation for spin (\ref{spinevolution}) reads
\begin{gather}
-i\eta \omega \frac{d \chi'}{d\theta} = \left[ \beta_0 \cdot \sigma -  \omega \delta + i \eta \omega g^{-1} \frac{ \partial g}{\partial \theta} \right] \chi' \ \ , \nonumber \\
 = \left[ ( \beta_0 + \frac{ \eta \omega W \hat{z}}{2})\cdot \sigma - \omega \delta \right] \chi' \ \ .
 \label{corot}
\end{gather}
The effective magnetic field in the co-rotating frame is static,
\begin{gather}
\mathcal{B} =\beta_0 + \frac{\eta \omega W}{2} \hat{z} \ \  ,
\label{B}
\end{gather}
and direct integration gives the evolution operator (in the laboratory frame)
\begin{gather}
U(\theta) = e^{- \frac{ i W \sigma_z \theta}{2}}e^{\frac{ i\eta \theta}{\omega} \mathcal{B} \cdot \sigma}
\end{gather}
and the eigenvalues of $U(2\pi)$ are given by
\begin{gather}
e^{\pm i \Phi} = e^{ -i \pi W \pm \frac{ 2\pi i \eta |\mathcal{B}|}{\omega} } \ \ .
\end{gather}
The phase $\Phi$ of the eigenvalues are unambiguously defined only up to a multiple of $2\pi$. In order to select the phase $\Phi$, we note that the Born-Oppenheimer approximation is valid only when the spin state $\chi$ is slowly varying. The spin states (in the laboratory frame) are given explicitly by
\begin{gather}
\chi_+(\theta) = 
e^{\frac{ i\eta  |\mathcal{B}| \theta}{\omega} -\frac{ i \Phi \theta}{2\pi}}( \cos \frac{\zeta}{2} e^{- \frac{ i W \theta }{2}}|+\rangle + \sin \frac{\zeta}{2} e^{\frac{ i W\theta}{2} + i \phi} |- \rangle )
\ \ \ , \nonumber \\
\chi_-(\theta) = 
e^{- \frac{i |\mathcal{B}|\theta }{\omega}  + \frac{i \Phi \theta}{2\pi} }
( -\sin \frac{\zeta}{2} e^{- \frac{ i W \theta}{2} - i \phi}|+\rangle + \cos \frac{\zeta}{2} e^{\frac{ i W \theta}{2}} |- \rangle )
\label{prec}
\end{gather}
where $\zeta$ is the angle between $\mathcal{B}$ and the plane,
\begin{gather}
\tan \zeta = \frac{\mathcal{B}_\parallel(\nu)}{\mathcal{B}_z(\nu)} \ \ ,
\label{zeta}
\end{gather}
$|+\rangle, |-\rangle$ are spin states with polarization along the $z$-axis, and  $\beta_0 = (\beta_\parallel \cos  \phi, \beta_\parallel \sin \phi, \beta_z)$.  The spin-up and spin-down components of $\chi_\pm$ accumulate different phases over the orbital trajectory. We may define $\Phi$ so that the largest spin component of $\chi_+$ is constant, with the smaller spin component acquiring a phase of $2\pi W$ around the trajectory. This choice ensures that as the spin-orbit interaction is reduced to zero, the energies $E_{n\pm}$ and quantum states $\chi_\pm$ are continuously related to the simply Zeeman-split levels in a uniform magnetic field. For the case $\mathcal{B}_z > 0$, the spin polarization is tilted above the plane and the largest spin component is $\langle + | \chi_+ \rangle$, while for $\mathcal{B}_z < 0$ the spin polarization is tilted below the plane and the largest component is $\langle - | \chi_+\rangle$, thus:\cblack
\begin{gather}
\Phi =  \frac{2 \pi \eta |\mathcal{B}|}{\omega}  - \pi W \text{sgn}(\beta_z +\frac{\eta \omega W}{2})  \ \ .
\label{PhiRot}
\end{gather}
Thus the wavefunctions are given by
\begin{gather}
\psi_{n+}(\theta) =
e^{- i \eta n\theta} (\cos \frac{\zeta_n}{2} |+\rangle + \sin \frac{\zeta_n}{2} e^{i W \theta + i \phi} |-\rangle) 
 \ \ , \nonumber \\
\psi_{n-}(\theta)
= e^{-i \eta n \theta} ( - \sin \frac{\zeta_n}{2} e^{-i W \theta - i \phi} |+\rangle 
+ \cos \frac{\zeta_n}{2} |-\rangle) 
\label{rotwavef1}
\end{gather}
for $\mathcal{B}_z = \beta_z + \frac{ \eta \omega W}{2} > 0$, and
\begin{gather}
\psi_{n+}(\theta) =
e^{-i \eta n \theta} ( \cos \frac{\zeta_n}{2} e^{- i W \theta -i \phi} |+\rangle
+ \sin \frac{\zeta_n}{2} |-\rangle)
 \ \ , \nonumber \\
\psi_{n-}(\theta)
e^{-i\eta n \theta}( - \sin \frac{\zeta_n}{2} |+\rangle + \cos \frac{\zeta_n}{2} e^{i W \theta + i \phi} |-\rangle) 
\label{rotwavef2}
\end{gather}
for $\mathcal{B}_z = \beta_z + \frac{ \eta \omega W}{2} < 0$, and   $\zeta_n$ correspond to angles $\zeta$ (\ref{zeta}) evaluated for values of $\nu = n$. The spin polarization in the upper spin state $\psi_+$ is along $S \parallel \beta + \frac{ \eta \omega W \hat{z}}{2}$, which is tilted out of the plane due to the rotation of the effective magnetic field.  This out-of-plane tilting is due to a geometric term $ i \eta  g^{-1} \frac{ \partial g}{\partial \theta}$ in the equation of motion (\ref{corot}). In the adiabatic limit $\omega \ll |\beta|$, spin will align along the direction of the effective magnetic field, $S \parallel \beta$, nevertheless the geometric contribution leads to a correction to $\Phi$ which is equal to the Berry phase discussed in Section IIIA.

It follows from (\ref{spectrum}) that the spectrum is
\begin{gather}
E_{n, \pm} = \omega(n+ \frac{1}{2}) \pm \left[ |\mathcal{B}| - \frac{\eta \omega W}{2} \text{sgn} ( \beta_z + \frac{\eta \omega W}{2}) \right] \nonumber \\
= \omega(n + \frac{1}{2} ) \nonumber \\
\pm \left[ \sqrt{( \beta_z + \frac{\eta \omega W}{2})^2 + \beta_\parallel^2} - \frac{\eta \omega W}{2} \text{sgn}( \beta_z + \frac{\eta \omega W}{2} )\right]
\label{spectrumnonabelian}
\end{gather}
 where the effective magnetic field $\beta(\nu)$ is taken along momentum space orbits corresponding to integer values of $\nu$.   We note that, while the choice of phase (\ref{PhiRot}) minimises the error in the semiclassical solution, we may arbitrarily redefine the phase by addition of an integer multiple of $2\pi$. After re-labeling of Landau levels (which only affects the ground state), addition of $2\pi$ to the phase is equivalent to a shift of index in the spin-dependent part of the energy, $E_{n, \pm} = \omega(n + \frac{1}{2} + \frac{\Phi_n}{2\pi}) \rightarrow \omega(n + \frac{1}{2} + \frac{ \Phi_{n+1}}{2\pi})$, which leads to an error of the same order in the semiclassical limit, although the numerical error may be larger for alternative choices of $\Phi$. We discuss this point further in Appendix A.  In the remainder of the section we shall apply these results to specific cases and compare them to the exact solutions.

\subsection{Rashba interaction in $n$-type systems}

The case of Rashba and Dresselhaus interactions in $n$-type systems has been extensively discussed in previous literatue.\cite{ZareaUlloa,ZhangLandau,PappMicu,Erlingsson}; we will review only the situation in which one of these interactions is present.
For the Rashba interaction, the Hamiltonian is given by
\begin{gather}
H = \frac{\bm{\pi}^2}{2m} + \alpha_R ( \bm{\pi}_y \sigma_x - \bm{\pi}_x \sigma_y) - \frac{g \mu_B B_z}{2}\sigma_z \ \ ,
\label{HRashba}
\end{gather}
where $\alpha_R$ is the Rashba constant, and the effective magnetic field  $\beta(\pi_x, \pi_y) = ( \alpha_R \pi_y, -\alpha_R \pi_x, -\frac{g \mu_B B_z}{2})$ has winding number $W = +1$. From (\ref{spectrumnonabelian}) the semiclassical solution is given by
\begin{widetext}

\begin{gather}
E_{n, \pm} = \omega(n + \frac{1}{2})  \pm
\left[ \sqrt{( - \frac{\omega}{2} - \frac{g \mu_B B_z}{2})^2 + \alpha_R^2 |2e B_z|(n+ \frac{1}{2})} + \frac{\omega}{2} \text{sgn}( -\frac{\omega}{2} - \frac{g \mu_B B_z}{2})\right] \ \ .
\label{spectrumRashbaSC}
\end{gather}
\cblack
A derivation of the exact solution is presented in Appendix B. We obtain the exact spectrum
\begin{gather}
E_{n,+} = \begin{cases}
\omega(n + \frac{1}{2}) + \left[ \sqrt{( \frac{\omega}{2} + \frac{g \mu_B B_z}{2})^2 + |2e B_z|\alpha_R^2 (n+1)} + \frac{\omega}{2}\right] \ \ , \ \  - \frac{\omega}{2} - \frac{g \mu_B B_z}{2} > 0 \ \ , \\
\omega (n + \frac{1}{2}) +\left[\sqrt{( \frac{\omega}{2} + \frac{g \mu_B B_z}{2})^2 + |2e B_z| \alpha_R^2 n} - \frac{\omega}{2} \right]\ \ , \ \ - \frac{\omega}{2} - \frac{g \mu_B B_z}{2} < 0
\end{cases} \ \ , \nonumber \\
E_{n,-} = \begin{cases}
\omega(n + \frac{1}{2}) - \left[ \sqrt{( \frac{\omega}{2} + \frac{g \mu_B B_z}{2})^2 + |2e B_z|\alpha_R^2 n} + \frac{\omega}{2}\right] \ \ , \ \ - \frac{\omega}{2} - \frac{ g \mu_B B_z}{2} > 0 \ \ , \\
\omega(n + \frac{1}{2}) - \left[ \sqrt{ ( \frac{\omega}{2} + \frac{g \mu_B B_z}{2})^2 + |2e B_z|\alpha_R^2 (n+1) } - \frac{\omega}{2}\right] \ \ , \ \  - \frac{\omega}{2} - \frac{g \mu_B B_z}{2} < 0
\end{cases}
\label{spectrumRashba}
\end{gather}
\end{widetext}
The exact wavefunctions are given by
\begin{gather}
\psi_{n,+}(\theta) = e^{i n \theta} ( \cos \frac{\zeta_{n+1}}{2} |+ \rangle +i \sin \frac{ \zeta_{n+1}}{2} e^{i \theta} |-\rangle) \ \ , \nonumber \\
\psi_{n,-}(\theta) = e^{i n\theta} ( -\sin \frac{ \zeta_n}{2} e^{-i \theta} |+ \rangle +i \cos \frac{\zeta_n}{2} |-\rangle)
\label{Rashbawave1}
\end{gather}
for $- \frac{ \omega}{2} - \frac{ g \mu_B B_z}{2} > 0$, and
\begin{gather}
\psi_{n,+}(\theta) = e^{i n \theta} ( \cos \frac{ \zeta_n}{2} e^{-i \theta} |+ \rangle + i\sin \frac{\zeta_n}{2} |-\rangle) \ \ , \nonumber \\
\psi_{n,-}(\theta) = e^{i n \theta} ( - \sin \frac{\zeta_{n+1}}{2} |+\rangle + i \cos \frac{\zeta_{n+1}}{2} e^{i \theta} |-\rangle)
\label{Rashbawave2}
\end{gather}
for $-\frac{ \omega}{2} - \frac{g \mu_B B_z}{2} < 0$. The angles $\zeta_n$ are defined in the same way as in the previous section (\ref{zeta}), $\tan \zeta_n = \frac{\alpha_R \sqrt{| 2e B_z| n}}{- \frac{ \omega}{2} - \frac{g \mu_B B_z}{2}}$.

~\\

\begin{figure}[h]
	 \includegraphics[width = 0.47\textwidth]{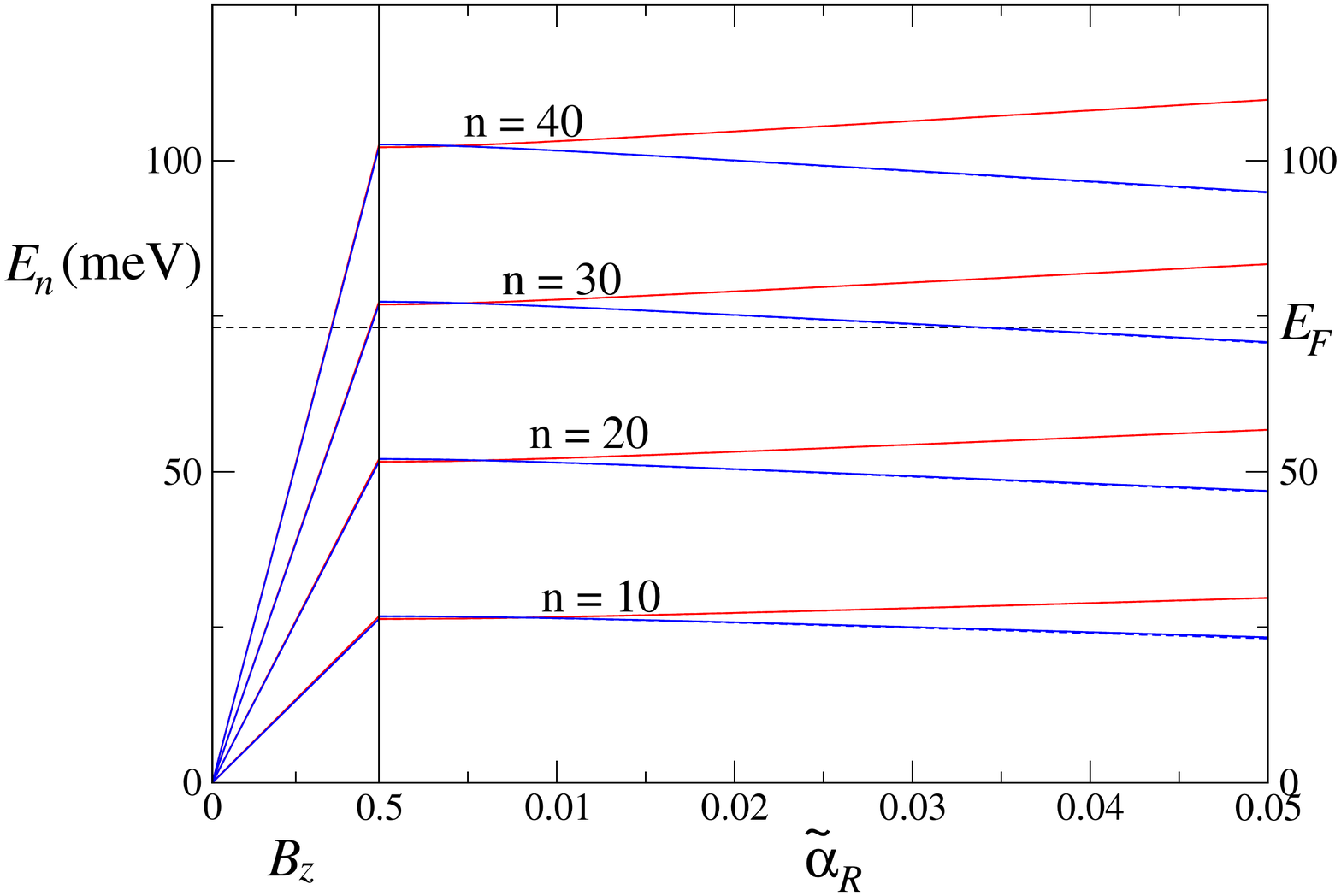}
	 \caption{
(\emph{Color online}). Energies of Landau level states $n = 10, 20, 30, 40$ in $n$-type 2D system in the presence of a Rashba interaction, plotted as a function of the dimensionless constant $\tilde{\alpha}_R =   \frac{ \alpha_R p_F}{E_F}$ where $E_F = 73\text{meV}$ is the Fermi energy corresponding to a 2D electron gas with typical experimental density\cite{Engels} $\rho = 7 \times 10^{11} \text{cm}^{-2}$ (the Fermi energy is indicated by the dashed horizontal line) and band parameters corresponding to InAs. The left panel  shows energies as a function of $B_z$ at $\alpha_R = 0$ and the right panel shows energies as a function of $\tilde{\alpha}_R$ at $B_z = 0.5\text{T}$. Red and blue lines indicate states of opposite spin. The difference between the exact  (\ref{spectrumRashba}) and semiclassical (\ref{spectrumRashbaSC}) solutions is not visible.}
\end{figure}

~\\

\begin{figure}[h]
	\includegraphics[width = 0.5\textwidth]{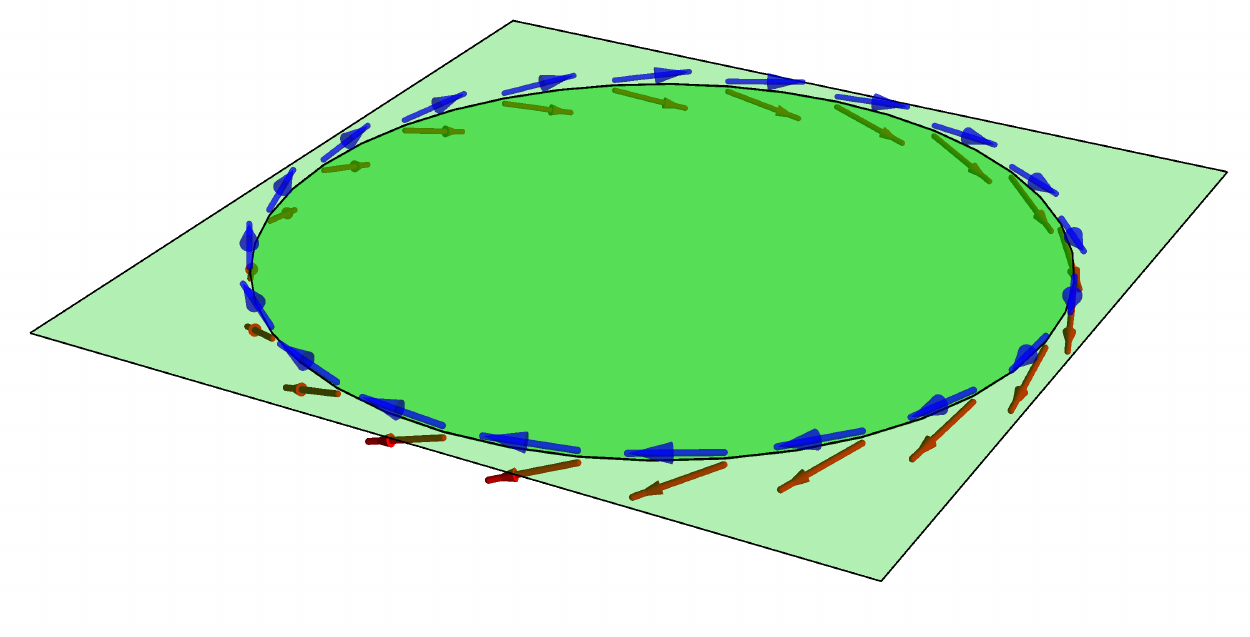}
	\caption{(\emph{Color online.}) Spin precession of Landau eigenstates along the momentum space trajectory due to the Rashba interaction in $n$-type InAs, shown for the highest filled Landau level ($n = 29$) at experimental density \cite{Engels} $\rho = 7 \times 10^{11} \text{cm}^{-2}$. The spin polarization is indicated by red arrows, and the effective magnetic field $\beta$ is indicated by blue arrows.}
\end{figure}

~\\

The error in the semiclassical solution is   $\approx \frac{\sin^2 \zeta}{8n} (E_{n+} - E_{n-})$. \cblack The Landau level energies for levels $n = 10, 20, 30, 40$ at $B_z = 0.5\text{T}$ are plotted in Fig. 1 as a function of the dimensionless parameter $\tilde{\alpha}_R = \frac{ \alpha_R p_F}{E_F}$ where $E_F = 73 \text{meV}, p_F$ are the Fermi energy and momentum corresponding to a 2D electron gas at experimental density\cite{Engels} $\rho = 0.6 \times 10^{12} \text{cm}^{-2}$. The band parameters are taken for InAs\cite{WinklerBook}, $m = 0.0229m_e, g = -14.9$. The semiclassical and exact results are both shown, although in this situation they are indistinguishable. For $n \gg 1$ the wavefunctions (\ref{Rashbawave1}) and (\ref{Rashbawave2}) reduce to the semiclassical expressions (\ref{rotwavef1}) and (\ref{rotwavef2}) with $W = +1$ and $\phi = \frac{ \pi}{2}$.

The precessing wavefunction is illustrated in Fig. 2 for the highest filled Landau level ($n = 29$) at the experimental density with the same parameters used in Fig. 1. The spin polarization $\psi^\dagger(\theta) \sigma \psi(\theta)$ is indicated by red arrows and the effective magnetic field $\beta(\theta)$ is indicated by blue arrows. While the effective magnetic field is tilted above the plane, the spin polarization is tilted below the plane, illustrating the size of the geometric contribution (\ref{B}) $\mathcal{B} - \beta_0 = -\frac{ \omega}{2} \hat{z}$  in the experimental parameter r\'{e}gime.

\subsection{Dresselhaus interaction in $n$-type systems}

We consider only the linear Dresselhaus interaction in (100)-oriented heterostructures, for which the  Hamiltonian is given by
\begin{gather}
H = \frac{ \bm{\pi}^ 2}{2m} + \alpha_D (\bm{\pi}_x \sigma_x - \bm{\pi}_y \sigma_y) - \frac{ g \mu_B B_z}{2} \sigma_z \  \ ,
\end{gather}
where $ \alpha_D$ is the Dresselhaus constant for the heterostructure. The rotating effective magnetic field, $\beta(\pi_x, \pi_y) = ( \alpha_D \pi_x, -\alpha_D \pi_y, -\frac{ g \mu_B B_z}{2})$ has winding number $W = -1$, and the semiclassical solution is identical in form to the solution for the Rashba case (\ref{spectrumRashba}) with the exception that, in the spin-dependent part $\omega$ is replaced by $-\omega$ due to the opposite winding number:
\begin{widetext}
\begin{gather}
E_{n, \pm} = \omega(n + \frac{1}{2}) \pm \left[ \sqrt{( \frac{\omega}{2} - \frac{g \mu_B B_z}{2})^2 + \alpha_D^2 |2e B_z|n} - \frac{\omega}{2} \text{sgn} ( \frac{\omega}{2} - \frac{g \mu_B B_z}{2})\right] \ \ .
\end{gather}
\cblack
A derivation of the exact solution is presented in Appendix B. The exact energies are given by
	\begin{gather}	\label{dressexact}
	E_{n,+} = 
	\begin{cases}
	\omega (n + \frac{1}{2}) + \left[ \sqrt{( \frac{\omega}{2} - \frac{g\mu_B B_z}{2})^2 + |2e B_z| \alpha_D^2 n } - \frac{ \omega}{2} \right] \ \ , \ \ \frac{ \omega}{2} - \frac{g \mu_B B_z}{2} > 0 \ \ , \\
	\omega(n + \frac{1}{2} + \left[ \sqrt{( \frac{\omega}{2} - \frac{g \mu_B B_z}{2})^2 + |2e B_z|\alpha_D^2 (n+1)} + \frac{ \omega}{2} \right] \ \ , \ \ \frac{\omega}{2} - \frac{g \mu_B B_z}{2} < 0 \ \ .
	\end{cases} \\ \nonumber
	E_{n,-}
	= \begin{cases}
	\omega (n + \frac{1}{2}) - \left[\sqrt{ ( \frac{\omega}{2} - \frac{g \mu_B B_z}{2})^2 + |2e B_z| \alpha_D^2 (n+ 1)} - \frac{\omega}{2} \right] \ \ , \ \ \frac{\omega}{2} - \frac{g \mu_B B_z}{2} \ \ , \nonumber \\
	\omega(n + \frac{1}{2}) - \left[ \sqrt{( \frac{ \omega}{2} - \frac{g \mu_B B_z}{2})^2 + |2e B_z| \alpha_D^2 n} + \frac{ \omega}{2} \right] \ \ , \ \ \frac{\omega}{2} - \frac{g \mu_B B_z}{2} < 0 \ \ .
	\end{cases}
	\end{gather}
\end{widetext}
The exact wavefunctions are given by
\begin{gather}
\psi_{n,+}(\theta) = e^{i n \theta} ( \cos \frac{\zeta_n}{2} | + \rangle + \sin \frac{ \zeta_n}{2} e^{ -i \theta} |- \rangle) \ \ , \nonumber \\
\psi_{n,-}(\theta)= e^{i n \theta} ( - \sin \frac{\zeta_{n+1}}{2} e^{i \theta} |+ \rangle + \cos \frac{ \zeta_{n+1}}{2} | -\rangle )
\label{wavedress1}
\end{gather}
for $ \frac{ \omega}{2} - \frac{g \mu_B B_z}{2} > 0$, and
\begin{gather}
\psi_{n,+}(\theta) = e^{i n \theta} ( \cos \frac{\zeta_{n+1}}{2} e^{i \theta} |+ \rangle + \sin \frac{ \zeta_{n+1}}{2} |-\rangle) \ \ , \nonumber \\
\psi_{n,-}(\theta) = e^{i n \theta} (- \sin \frac{\zeta_n}{2} |+\rangle + \cos \frac{ \zeta_n}{2} e^{ -i \theta} |-\rangle)
\label{wavedress2}
\end{gather}
for $\frac{\omega}{2} - \frac{g \mu_B B_z}{2} < 0$. For $n \gg 1$, the wavefunctions reduce to their semiclassical expressions (\ref{rotwavef1}), (\ref{rotwavef2}) with $W = -1$ and $\phi = 0$.

\subsection{$p$-type systems with in-plane magnetic field}

The Hamiltonian in case when both in-plane and perpendicular components of the magnetic field are present is given by
\begin{gather}
H = \frac{\bm{\pi}^2}{2m} - \frac{\alpha_H}{2} (B_+ \bm{\pi}_+^2 \sigma_- + B_- \bm{\pi}_-^2 \sigma_+ ) - \frac{ g \mu_B B_z}{2} \ \ .
\end{gather}
where $\alpha_H$ is a constant which depends on the 2D confining potential and the bulk $g$-factor. The rotating effective magnetic field is $\beta(\pi_x, \pi_y) = -(\alpha_H B_\parallel \pi_\parallel^2 \cos (2\theta + \phi), \alpha_H B_\parallel \pi_\parallel^2 \sin (2\theta+ \phi), \frac{ g \mu_B B_z}{2})$ where $B_+ = B_\parallel e^{i \phi}$ and has winding number $W = 2$. Thus  the semiclassical spectrum (\ref{spectrumnonabelian}) is given by

\begin{figure}[h]
	 \includegraphics[width = 0.47\textwidth]{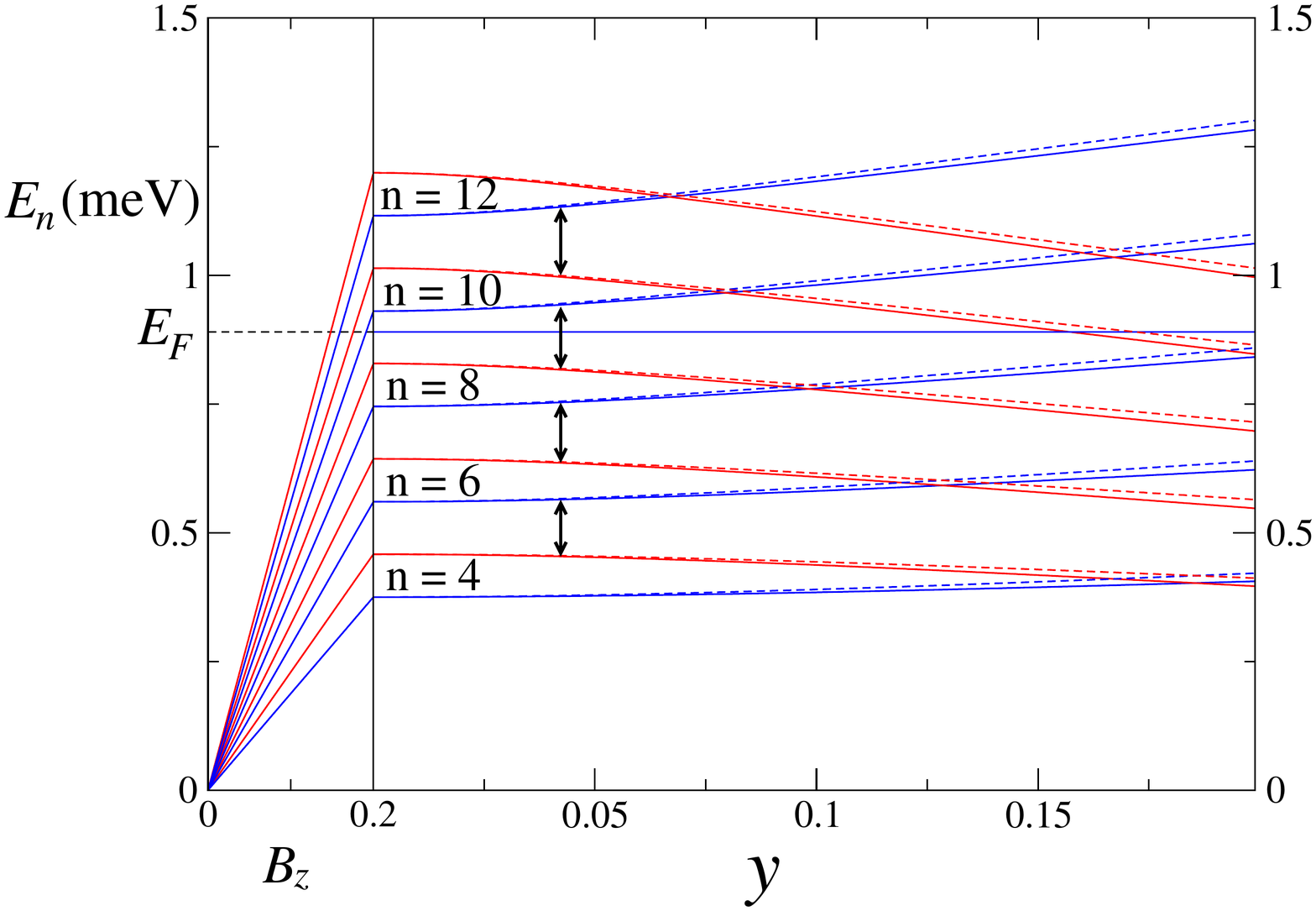}
	 \caption{
(\emph{Color online}). Energies of Landau level states $n = 4, 6, 8, 10, 12$ in a 2D GaAs hole gas in the presence of an in-plane magnetic field $B_\parallel$, plotted as a function of the dimensionless constant $y = 2m \alpha_H B_\parallel$. The left panel  shows energies as a function of $B_z$ at $y = 0$ and the right panel shows energies as a function of $y$ at $B_z = 0.2\text{T}$. The exact solutions 
(51)
are indicated in solid lines, and the semiclassical approximation
(50)
is indicated in dashed lines. Red and blue lines indicate states of opposite spin. The Fermi energy $E_F = 0.89\text{meV}$ corresponding to the typical experimental density\cite{LiYeoh} $\rho = 9.3 \times 10^{10}\text{cm}^{-2}$ is indicated by the dashed horizontal line. The arrows indicate possible ESR transitions (discussed in Section V).
}
\end{figure}

\begin{widetext}

\begin{gather}
E_{n,\pm} = \omega (n + \frac{1}{2}) \pm
 \left[ \sqrt{(\omega - \frac{g\mu_B B_z}{2})^2 + (2e \alpha_H B_\parallel B_z)^2 (n + \frac{1}{2})^2} - \omega \text{sgn}( \omega - \frac{g \mu_B B_z}{2}) \right] \ \ .
 \label{inplanepSC}
\end{gather}
\cblack
A derivation of the exact solution is presented in Appendix C. The exact energies are given by
	\begin{gather} \label{inplanepexact}
	E_{n+ } = \begin{cases}
	\omega( n + \frac{1}{2}) + \left[ \sqrt{( \omega - \frac{ g \mu_B B_z}{2})^2 + (2e \alpha_H B_z B_\parallel \nu_n)^2 } - \omega \right] \ \ , \ \ \omega - \frac{g \mu_B B_z}{2} > 0 \ \ , \\
	\omega( n + \frac{1}{2}) + \left[ \sqrt{( \omega - \frac{g \mu_B B_z}{2})^2 + (2 e \alpha_H B_z B_\parallel \nu_{n+2})^2} + \omega \right] \ \ , \ \ \omega - \frac{g \mu_B B_z}{2} < 0 \ \ .
	\end{cases} \\ \nonumber 
	E_{n-} = \begin{cases}
	\omega(n + \frac{1}{2}) - \left[ \sqrt{( \omega - \frac{g \mu_B B_z}{2})^2 + (2e \alpha_H B_z B_\parallel \nu_{n+2})^2} - \omega \right] \ \ , \ \ \omega - \frac{g \mu_B B_z}{2} > 0 \ \ , \\
	\omega( n + \frac{1}{2}) - \left[ \sqrt{( \omega - \frac{g \mu_B B_z}{2})^2 + (2e\alpha_H B_z B_\parallel \nu_n)^2} + \omega \right] \ \ , \ \  \omega - \frac{g \mu_B B_z}{2} < 0 \ \ .
	\end{cases}
	\end{gather}
\end{widetext}
where
\begin{gather}
\nu_n = \sqrt{n(n-1)} \ \ .
\end{gather}
The exact wavefunctions are
\begin{gather} 
\psi_{n,+}(\theta) = e^{-i n \theta} ( \cos \frac{\zeta_{\nu_n}}{2} |+ \rangle + \sin \frac{ \zeta_{\nu_n}}{2} e^{2i \theta} |-\rangle) \ \ , \nonumber \\
\psi_{n,-}(\theta) = e^{-i n \theta} ( - \sin \frac{\zeta_{\nu_{n+2}}}{2} e^{- 2i \theta} |+\rangle + \cos \frac{\zeta_{\nu_{n+2}}}{2} |-\rangle) 
\label{inplanewave1}
\end{gather}
for $ \omega - \frac{g \mu_B B_z}{2} > 0$, with $B_+ = B_\parallel e^{i \phi}$ and
\begin{gather} 
\psi_{n,+}(\theta) = e^{-i n \theta} ( \cos \frac{\zeta_{\nu_{n+2}}}{2} e^{-2i \theta} |+ \rangle + \sin \frac{ \zeta_{\nu_{n+2}}}{2}  |-\rangle) \ \ , \nonumber \\
\psi_{n,-}(\theta) = e^{-i n \theta} ( - \sin \frac{\zeta_{\nu_{n} }}{2} |+\rangle + \cos \frac{\zeta_{\nu_n}}{2} e^{2i \theta} |-\rangle) 
\label{inplanewave2}
\end{gather}
for $\omega - \frac{g \mu_B B_z}{2} < 0$, and the angles $\zeta_{\nu_n}$ are given by
\begin{gather}
\tan \zeta_{\nu_n} = \frac{2e \alpha_H B_z B_\parallel \nu_n}{ \omega - \frac{g \mu_B B_z}{2}} \ \ .
\end{gather}
For $n \gg 1$ we have $\nu_n \rightarrow n$ and the exact wavefunctions reduce to the semiclassical expressions (\ref{rotwavef1}), (\ref{rotwavef2}) with $W = +2$.

The error in the semiclassical solution is  $\approx \frac{1}{2n} \sin^2 \zeta_{\nu_n} (E_{n+} - E_{n-})$. \cblack The Landau level energies for $n = 4,6,8,10,12$ are plotted in Fig. 3 as a function of the dimensionless parameter $y = 2m \alpha_H B_\parallel$ for a GaAs 2D hole system, with effective mass $m = 0.25m_e$ corresponding to the experimental situation reported in\cite{LiYeoh}. We also take a value for the $g$-factor in GaAs\cite{WinklerBook} $g = 6\kappa = 7.2$. \cblack The experimentally measured value of $\alpha_H$ in experiment corresponds to $y = 0.029$ at $B_\parallel = 1\text{T}$.  The exact solution (\ref{inplanepexact}) is shown in solid lines and the semiclassical solution (\ref{inplanepSC}) is shown in dashed lines. The horizontal dashed line indicates the Fermi energy at experimental density\cite{LiYeoh} $\rho = 9.3 \times 10^{10} \text{cm}^{-2}$.
~\\~\\

\begin{figure}[h]
\includegraphics[width = 0.47\textwidth]{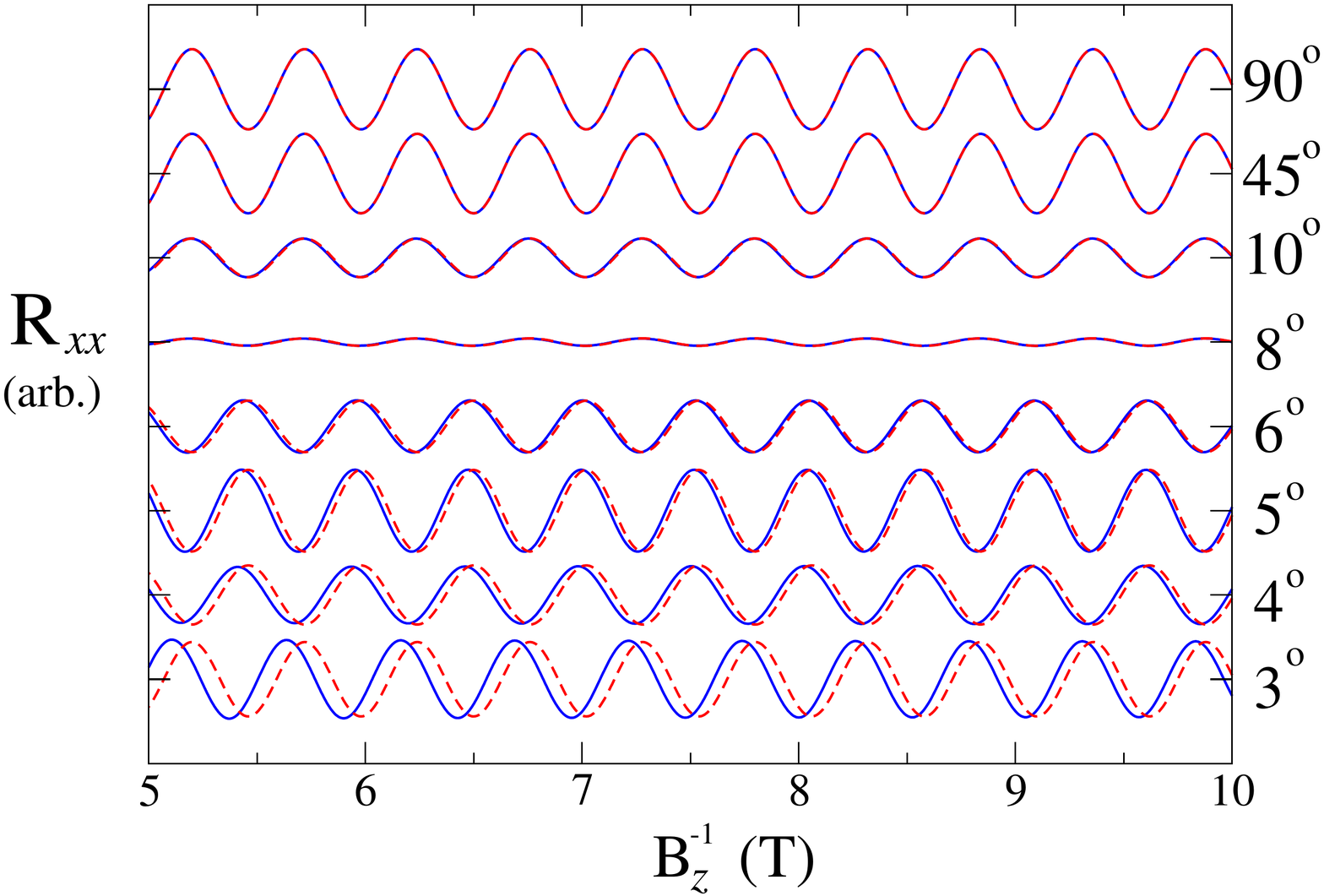}
\caption{
(\emph{Color online}). The oscillating resistivity $R_{xx}(B_z)$ (arbitrary units) as a function of $B_z^{-1}$, with the ratio $\frac{B_z}{B_x} = \tan \theta_{tilt}$ kept fixed. The values of $\theta_{tilt}$ corresponding to the individual traces are shown on the right of the figure. The solid lines indicate the oscillations obtained from the exact solution (51) while the dashed lines indicate the semiclassical solution (50). The semiclassical and exact results can only be distinguished for angles $\theta_{tilt} = 3^{\circ}, 4^{\circ}, 5^{\circ}$.
}
\end{figure}
~\\~\\

The oscillating resistivity $\frac{ R_{xx}(B_z)}{R_{xx}(B_z = 0)}$ is plotted in Fig. 4 for various values of $B_x$, with the ratio $\frac{B_z}{B_x} = \tan\theta_{tilt}$ kept fixed. The \emph{tilt angles} $\theta_{tilt}$ corresponding to the individual traces are shown on the right side of the figure. The solid line indicates the oscillations obtained from the exact solution (50), while the dashed line indicates the semiclassical solution (49). The semiclassical and exact results can only be distinguished at the lowest angles, $\theta_{tilt} = 3^{\circ}, 4^{\circ}, 5^{\circ}$. The precessing wavefunction is illustrated in Fig. 5 for the highest filled Landau level ($n = 9$ at $B_z = 0.2 \text{T}$) at the experimental density with the same band parameters used in Fig. 3. The in-plane magnetic field $B_x$ corresponds to a value $y = 2m \alpha_H B_x = 0.116$. The spin polarization $\psi^\dagger(\theta) \sigma \psi(\theta)$ is indicated by red arrows and the effective magnetic field $\beta(\theta)$ is indicated by blue arrows. The difference between the spin polarization and the effective magnetic field is given by the geometric contribution (\ref{B}) $\mathcal{B} - \beta_0 =  \omega \hat{z}$ corresponding to a rotating effective magnetic field with winding number $W = +2$.

~\\~\\

\begin{figure}[h]
	\includegraphics[width = 0.5\textwidth]{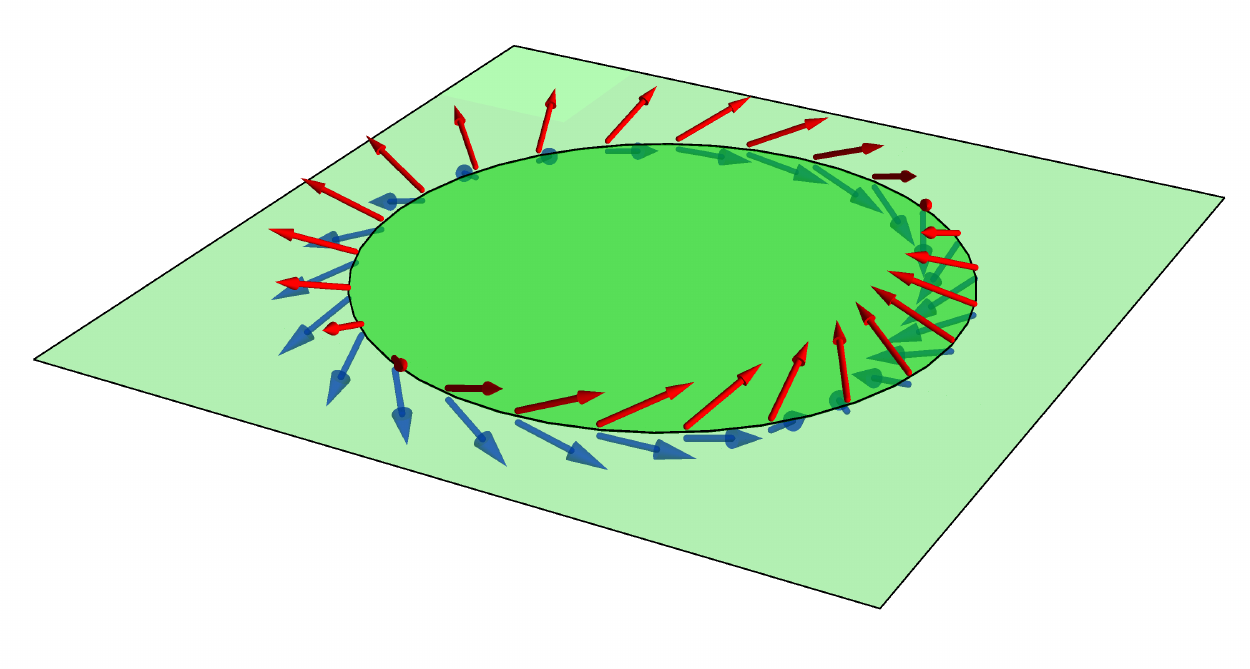}
	\caption{(\emph{Color online.}) Spin precession of a Landau level eigenstate along the momentum space trajectory due to an in-plane magnetic field 
$B_x$ corresponding to $y = 2m \alpha_H B_x = 0.116$  in a GaAs hole gas, shown for the highest filled Landau level ($n = 9$, $B_z = 0.2\text{T}$) at experimental density \cite{LiYeoh} $\rho = 9.3 \times 10^{10} \text{cm}^{-2}$. The spin polarization is indicated by red arrows, and the effective magnetic field $\beta$ is indicated by blue arrows.}
\end{figure}

\subsection{Rashba interaction in $p$-type systems}

The Hamiltonian in this case is given by
\begin{gather}
H = \frac{\bm{\pi}^2}{2m} + \frac{i\alpha'_R}{2} ( \bm{\pi}_+^3 \sigma_- - \bm{\pi}_-^3 \sigma_+)
\end{gather}
where $\alpha'_R$ is the Rashba constant for the heterostructure. The rotating effective magnetic field, $\beta(\pi_x, \pi_y) = ( \alpha'_R \pi^3_\parallel \sin 3\theta, -\alpha'_R \pi_\parallel^3 \cos 3\theta,  \frac{g \mu_B B_z}{2})$ has winding number $W = 3$. The semiclassical solution is given by (\ref{spectrumnonabelian})
\begin{widetext}

\begin{gather}
E_{n,\pm} = \omega(n + \frac{1}{2}) \pm
\left[ \sqrt{ ( \frac{3 \omega}{2} - \frac{g \mu_B B_z}{2})^2 + (\alpha'_R)^2 (2e B_z)^3 (n + \frac{1}{2})^3} - \frac{3 \omega}{2} \text{sgn}( \frac{3 \omega}{2} - \frac{g \mu_B B_z}{2}) \right] \ \ .
	\label{rashbapSC}
\end{gather}
\cblack
A derivation of the exact solution is presented in Appendix C. The exact energies are given by
	\begin{gather}
				\label{rashbapexact}
	E_{n,+} = \begin{cases}
	\omega( n + \frac{1}{2}) + \left[ \sqrt{( \frac{3 \omega}{2} - \frac{g \mu_B B_z}{2})^2 + (\alpha'_R)^2 (2e B_z \nu_n)^3 } - \frac{3 \omega}{2} \right] \ \ , \ \ \frac{3 \omega}{2} - \frac{g \mu_B B_z}{2} > 0 \ \ , \\
	\omega( n + \frac{1}{2}) + \left[ \sqrt{ ( \frac{3 \omega}{2} - \frac{g \mu_B B_z}{2})^2 + (\alpha'_R)^2 ( 2e B_z \nu_{n+3})^3 } + \frac{3 \omega}{2} \right] \ \ , \ \ \frac{3 \omega}{2} - \frac{g \mu_B B_z}{2} < 0
	\end{cases} \\ \nonumber
	E_{n,-} = \begin{cases}
	\omega(n + \frac{1}{2}) - \left[ \sqrt{( \frac{3 \omega}{2} - \frac{g \mu_B B_z}{2})^2 + (\alpha'_R)^2 ( 2e B_z \nu_{n+3})^2 } - \frac{3 \omega}{2} \right] \ \ , \ \ \frac{3 \omega}{2} - \frac{g \mu_B B_z}{2} > 0 \ \ , \\
	\omega(n + \frac{1}{2}) - \left[ \sqrt{( \frac{ 3 \omega}{2} - \frac{g \mu_B B_z}{2})^2 + (\alpha'_R)^2 ( 2e B_z \nu_n)^2} + \frac{3 \omega}{2} \right] \ \ , \ \ \frac{3 \omega}{2} - \frac{g \mu_B B_z}{2} < 0
	\end{cases}
	\end{gather}
\end{widetext}
where 
\begin{gather}
\nu_n = (n(n-1)(n-2))^\frac{1}{3}
\end{gather}
and the wavefunctions are given by
\begin{gather}
\psi_{n,+}(\theta) = e^{-i n \theta}( \cos \frac{\zeta_{\nu_n}}{2} |+ \rangle + i \sin \frac{ \zeta_{\nu_n}}{2} e^{3i \theta} |-\rangle) \ \ , \nonumber \\
\psi_{n,-}(\theta) = e^{-i n \theta}( - \sin \frac{\zeta_{\nu_{n+3}}}{2} e^{ -3i \theta}|+\rangle + i \cos \frac{\zeta_{\nu_{n+3}}}{2} |-\rangle )
\label{rashbapwave1}
\end{gather}
for $\frac{3 \omega}{2} - \frac{g \mu_B B_z}{2} > 0$, and
\begin{gather}
\psi_{n,+}(\theta) = e^{-i n \theta} ( \cos \frac{\zeta_{\nu_{n+3}}}{2} e^{-3i \theta} |+\rangle + i \sin \frac{\zeta_{\nu_{n+3}}}{2} |-\rangle) \ \ , \nonumber \\
\psi_{n, -}(\theta) = e^{-i n \theta}(- \sin \frac{ \zeta_{\nu_n}}{2} |+ \rangle + i \cos \frac{\zeta_{\nu_n}}{2} e^{3i \theta} |-\rangle)
\label{rashbapwave2}
\end{gather}
for $ \frac{3 \omega}{2} - \frac{g \mu_B B_z}{2} < 0$. Here the angles $\zeta_{\nu_n}$ are given by
\begin{gather}
\tan \zeta_{\nu_n} = \frac{ \alpha'_R(2e B_z \nu_n)^\frac{3}{2}}{\frac{3 \omega}{2} - \frac{g \mu_B B_z}{2}} \ \ .
\end{gather}
For $n \gg 1$ we have $\nu_n \rightarrow n$ and the exact wavefunctions reduce to the semiclassical expressions (\ref{rotwavef1}), (\ref{rotwavef2}) with $W = +3, \phi = \frac{ \pi}{2}$.

The error in the semiclassical solution is  $\approx \frac{9}{8n} \sin^2 \zeta_{\nu_n} (E_{n+} - E_{n-})$. \cblack The Landau level energies at $B_z = 0.5\text{T}$ are plotted in Fig. 6 for $n = 4, 8, 12, 16$  as a function of the dimensionless parameter $\tilde{\alpha}'_R = \frac{\alpha'_R p_F^3}{E_F}$ where the Fermi energy $E_F = 2\text{meV}$ (indicated by the dashed horizontal line) corresponds to the experimental density\cite{Grbic} $\rho = 3 \times 10^{11} \text{cm}^{-2}$. The exact solutions (\ref{rashbapexact}) are shown in solid lines, and the semiclassical solutions (\ref{rashbapSC}) are shown in dashed lines.

\begin{figure}[h]
	 \includegraphics[width = 0.47\textwidth]{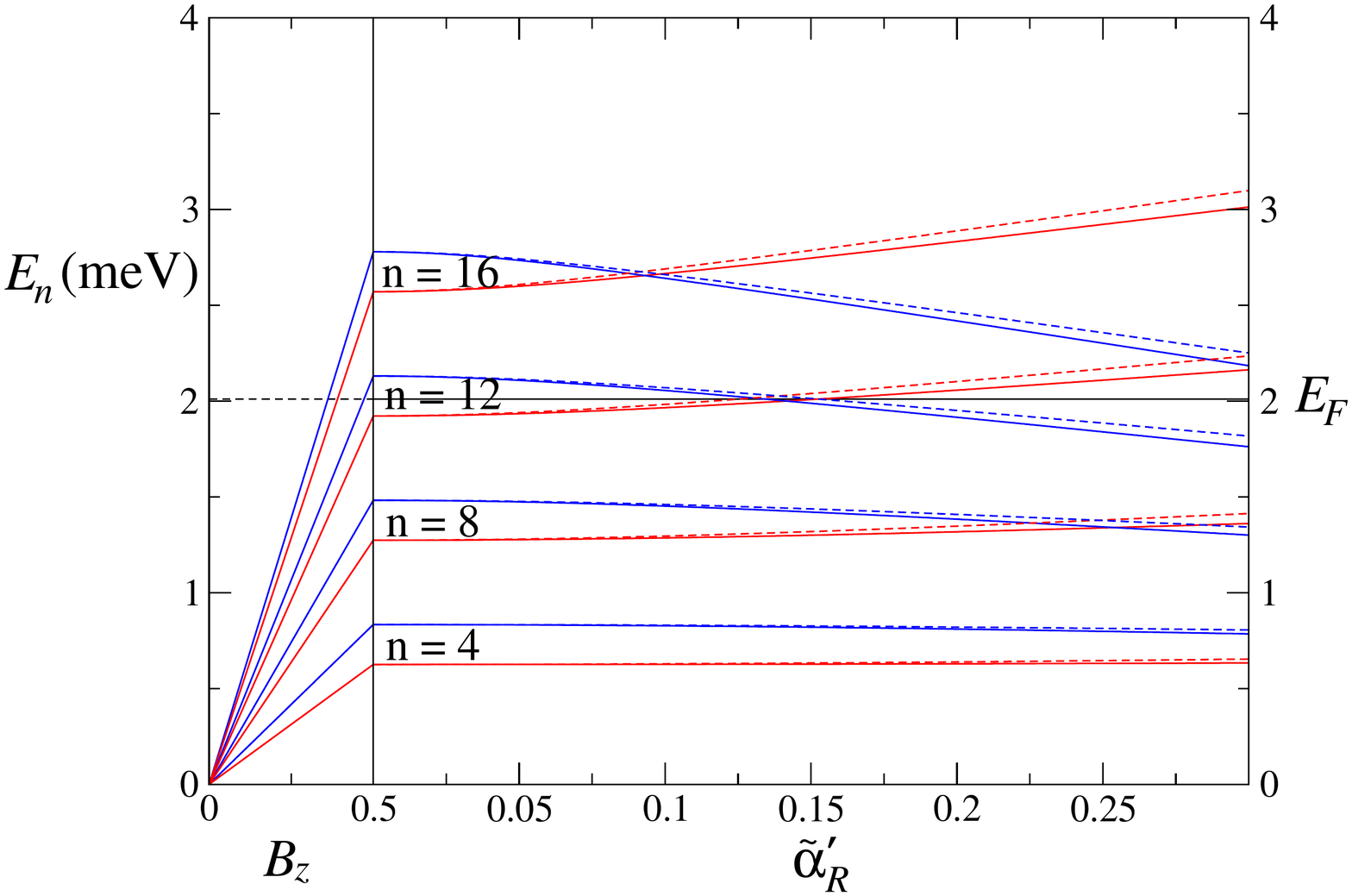}
	 \caption{
(\emph{Color online}). Energies of Landau level states $n = 4, 8, 12, 16$ in a 2D GaAs hole gas in the presence of a Rashba interaction, plotted as a function of the dimensionless constant $\tilde{\alpha}'_R = \frac{ \alpha_R p_F^3}{E_F}$ where the Fermi energy $E_F = 2\text{meV}$ (indicated by the dashed horizontal line) corresponds to the typical experimental density\cite{Grbic} $\rho = 3 \times 10^{11}\text{cm}^{-2}$. The left panel  shows energies as a function of $B_z$ at $\tilde{\alpha}'_R = 0$ and the right panel shows energies as a function of $\tilde{\alpha}'_R$ at $B_z = 0.5\text{T}$. The exact solutions (\ref{rashbapexact}) are indicated in solid lines, and the semiclassical approximation (\ref{rashbapSC}) is indicated in dashed lines. Red and blue lines indicate states of opposite spin.
}
\end{figure}

The precessing wavefunctions are illustrated in Fig. 6 for the highest filled Landau level ($n = 12$) at the experimental density with the same parameters used in Fig. 7. The spin polarization $\psi^\dagger(\theta) \sigma \psi(\theta)$ is indicated by red arrows and the effective magnetic field $\beta(\theta)$ is indicated by blue arrows. The difference between the spin polarization and the effective magnetic field is given by the geometric contribution (\ref{B}) $\mathcal{B} - \beta_0 =   \frac{3\omega}{2} \hat{z}$ corresponding to a rotating effective magnetic field with winding number $W = +3$.

\subsection{Dresselhaus interaction in $p$-type systems}

The Hamiltonian in the case of a pure Dresselhaus interaction is given by
\begin{gather}
H = \frac{ \bm{\pi}^2}{2m} - \frac{g \mu_B B_z}{2} \sigma_z \nonumber \\
+ \frac{\alpha'_D}{4} ( (\bm{\pi}_+^2 \bm{\pi}_- + \bm{\pi}_- \bm{\pi}_+^2)\sigma_- + (\bm{\pi}_+ \bm{\pi}_-^2 + \bm{\pi}_-^2 \bm{\pi}_+) \sigma_+)  \ \ ,
\end{gather}
where $\alpha'_D$ is the Dresselhaus constant for the heterostructure. The rotating effective magnetic field, $\beta(\pi_x, \pi_y) = ( -\alpha'_D \pi^2 \pi_y, \alpha'_D \pi^2 \pi_x, \frac{g \mu_B B_z}{2})$ has winding number $W = 1$. The semiclassical solution is therefore given by (\ref{spectrumnonabelian})

\begin{figure}[t]
	\includegraphics[width = 0.5\textwidth]{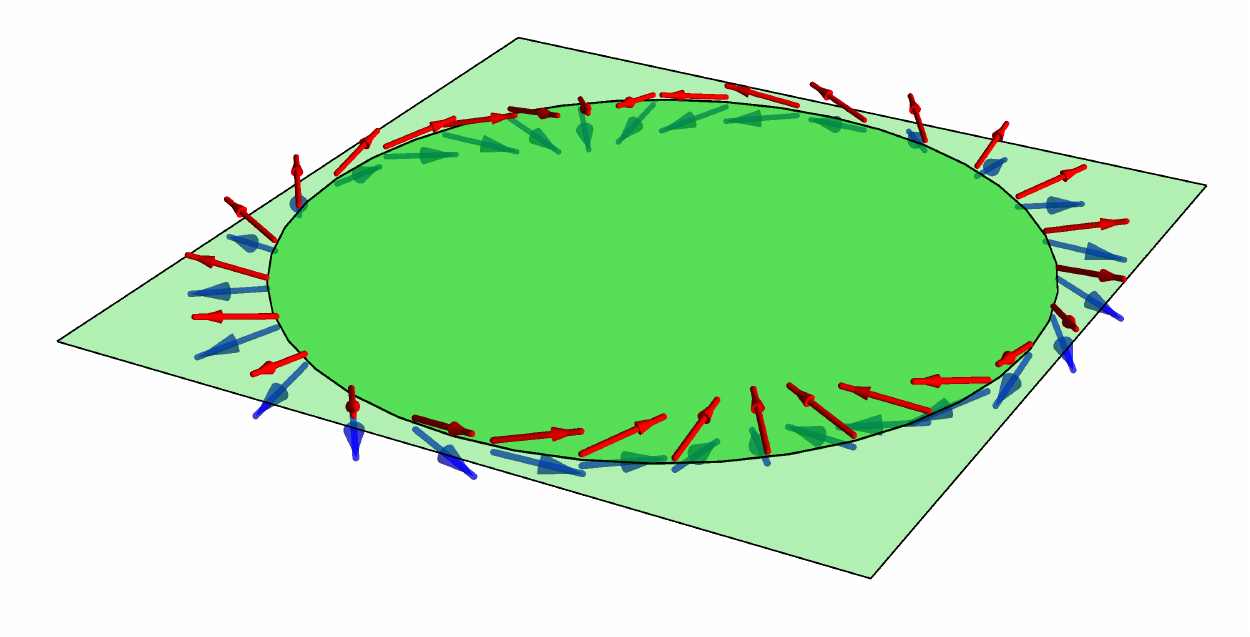}
	\caption{(\emph{Color online.}) Spin precession of Landau eigenstates along the momentum space trajectory due to the Rashba interaction shown for the highest filled Landau level ($n = 12$) at the experimental density\cite{Grbic} $\rho = 3\times 10^{11}\text{cm}^{-2}$. The spin polarization is indicated by red arrows, and the effective magnetic field $\beta$ is indicated by blue arrows.}
\end{figure}

\begin{widetext}

\begin{gather}
E_{n,\pm} = \omega(n + \frac{1}{2}) \pm
\left[ \sqrt{( \frac{\omega}{2} - \frac{g \mu_B B_z}{2})^2 + (\alpha'_D)^2 (2e B_z)^3 (n + \frac{1}{2})^3} - \frac{\omega}{2} \text{sgn}( \frac{ \omega}{2} - \frac{g \mu_B B_z}{2})\right] \ \ .
\end{gather}
\cblack
A derivation of the exact solution is presented in Appendix C. The exact energies are given by
	\begin{gather} \label{dresspexact}
	E_{n,+} = \begin{cases}
	\omega(n + \frac{1}{2}) + \left[ \sqrt{ ( \frac{ \omega}{2} - \frac{g \mu_B B_z}{2})^2 + (\alpha'_D)^2 (2e B_z n)^3} - \frac{\omega}{2} \right] \ \ , \ \ \frac{\omega}{2} - \frac{g \mu_B B_z}{2} > 0 \ \ , \\
	\omega(n+ \frac{1}{2}) + \left[ \sqrt{ ( \frac{\omega}{2} - \frac{g \mu_B B_z}{2})^2 + (\alpha'_D)^2 (2e B_z (n+1))^3} + \frac{ \omega}{2} \right] \ \ , \ \ \frac{\omega}{2} - \frac{g \mu_B B_z}{2} < 0
	\end{cases} \nonumber \\
	E_{n,-} = \begin{cases}
	\omega(n + \frac{1}{2}) - \left[ \sqrt{ ( \frac{ \omega}{2} -\frac{g \mu_B B_z}{2})^2 + (\alpha'_D)^2 (2e B_z)^2(n+1)^2} - \frac{\omega}{2} \right] \ \ , \ \ \frac{\omega}{2} - \frac{g \mu_B B_z}{2} > 0 \ \ , \\
	\omega(n + \frac{1}{2}) - \left[ \sqrt{(\frac{\omega}{2} - \frac{g \mu_B B_z}{2})^2 + (\alpha'_D)^2 (2e B_z n)^2} + \frac{\omega}{2} \right] \ \ , \ \ \frac{\omega}{2} - \frac{ g \mu_B B_z}{2} < 0
	\end{cases}
	\end{gather}
\end{widetext}
and the wavefunctions are given by
\begin{gather}
\psi_{n,+}(\theta) = e^{-i n \theta}( \cos \frac{\zeta_n}{2} |+\rangle + \sin \frac{\zeta_n}{2} e^{i \theta} |-\rangle) \nonumber \\
\psi_{n,-}(\theta) = e^{-i n \theta}( - \sin \frac{\zeta_{n+1}}{2} e^{-i \theta} |+\rangle + \cos \frac{ \zeta_{n+1}}{2} |-\rangle)
\label{dresspwave1}
\end{gather}
for $\frac{ \omega}{2} - \frac{g \mu_B B_z}{2} > 0$, and
\begin{gather}
\psi_{n,+}(\theta) = e^{-i n \theta}( \cos \frac{\zeta_{n+1}}{2} e^{-i \theta}|+\rangle + \sin \frac{\zeta_{n+1}}{2} |-\rangle) \nonumber \\
\psi_{n,-}(\theta) = e^{-i n \theta}( - \sin \frac{\zeta_n}{2}  |+\rangle + \cos \frac{ \zeta_{n}}{2} e^{i \theta}|-\rangle)
\label{dresspwave2}
\end{gather}
for $\frac{\omega}{2} - \frac{g \mu_B B_z}{2} < 0$. Here
\begin{gather}
\tan \zeta_n = \frac{ \alpha'_D (2e B_z n)^\frac{3}{2}}{ \frac{ \omega}{2} - \frac{g \mu_B B_z}{2}}  \ \ .
\end{gather}
For $n \gg 1$ the exact wavefunctions reduce to the semiclassical expressions (\ref{rotwavef1}), (\ref{rotwavef2}) with $W = +1, \phi = 0$.

\section{Electron spin resonance}
\begin{table*}[ht]
	\centering
	\begin{tabular}{|c|c |c|c|}
		\hline
		~
		& & Transition & $|\langle \beta | \delta H | \alpha\rangle|^2$  \\ \hline
		\multirow{12}{*}{Electrons} & \multirow{6}{*}{$-\frac{ W \omega}{2} - \frac{ g \mu_B B_z}{2} > 0 $}&
		$|n, + \rangle \rightarrow |n, - \rangle$ &
		$ \cos^4 \frac{\zeta}{2}$ \\ \cline{3-4}
		& &	$|n, + \rangle \rightarrow |n + 2 W, - \rangle$ & $ \sin^4 \frac{\zeta}{2} $ \\
		\cline{3-4}
		& &	$|n,+\rangle \rightarrow |n - W, + \rangle $ & \multirow{4}{*}{ $ \frac{1}{4}  \sin^2 \zeta$} \\ 
		& &$|n,+\rangle \rightarrow |n + W, + \rangle$
		& \\ 
		& &$|n, - \rangle \rightarrow |n - W, -\rangle$ & 
		\\ 
		& & $|n, - \rangle \rightarrow |n + W, - \rangle$ &
		\\
		\cline{2-4}
		& \multirow{6}{*}{$- \frac{ W \omega}{2} - \frac{ g \mu_B B_z}{2} < 0$} &
		$|n, + \rangle \rightarrow |n, -\rangle$ & $\sin^4\frac{\zeta}{2}$ \\ \cline{3-4}
		& & $|n, + \rangle \rightarrow |n-2W,-\rangle$ & $\cos^4\frac{\zeta}{2}$ \\ \cline{3-4}
		& & $|n, + \rangle \rightarrow |n+W,+\rangle$ & \multirow{4}{*} {$\frac{1}{4} \sin^2 \zeta$} \\
		& & $|n, - \rangle \rightarrow |n-W,+\rangle$ & \\
		& & $|n,-\rangle \rightarrow |n+W,-\rangle$ & \\
		& & $|n,-\rangle \rightarrow |n-W,-\rangle$ & \\
 \hline
		\multirow{14}{*}{Holes} & \multirow{7}{*}{$ \frac{ W \omega}{2} - \frac{ g \mu_B B_z}{2} > 0$}&
		$|n, + \rangle \rightarrow |n-2, -\rangle$ & $\cos^4 \frac{\zeta}{2}$ ($W \neq 2$) \\ \cline{3-4}
		& & $|n, + \rangle \rightarrow |n+2 - 2W, - \rangle$ & $\sin^4 \frac{\zeta}{2}$ ($W \neq 2$)\\ \cline{3-4}
		& & $|n, + \rangle \rightarrow |n-2, - \rangle$ & $1 - \frac{ \sin^2 \zeta}{2} \left[ 1 + \cos 2 (\phi - \varphi)\right]$ ($W = 2$) \\ \cline{3-4}
		& & $|n, + \rangle \rightarrow |n+2 - W, + \rangle$ & \multirow{5}{*}{$ \frac{1}{4} \sin^2 \zeta$} \\
		& & $|n, + \rangle \rightarrow |n+W-2, + \rangle$ & \\
		& & $|n, - \rangle \rightarrow |n+2 - W, - \rangle$ & \\
		& & $|n, - \rangle \rightarrow |n+W-2, - \rangle$ & \\ \cline{2-4}
		& \multirow{7}{*}{$\frac{ W \omega}{2} - \frac{ g \mu_B B_z}{2} < 0$} &
		$|n,+ \rangle \rightarrow |n+2, -\rangle$ & $\sin^4\frac{\zeta}{2}$ ($W \neq 2$) \\ \cline{3-4}
		& & $|n,+\rangle \rightarrow |n + 2W - 2,-\rangle$ & $\cos^4\frac{\zeta}{2}$ ($W \neq 2$) \\ \cline{3-4}
		& & $|n, + \rangle \rightarrow |n-2, - \rangle$ & $1 - \frac{ \sin^2 \zeta}{2} \left[ 1 + \cos 2 (\phi - \varphi)\right]$ ($W = 2$) \\ \cline{3-4}
		& & $|n, + \rangle \rightarrow |n +2 -W,+\rangle$ & \multirow{5}{*}{$ \frac{1}{4} \sin^2\zeta$ } \\
		& & $|n,+ \rangle  \rightarrow |n + W- 2,+ \rangle$ & \\
		& & $|n,-\rangle \rightarrow |n+2- W,-\rangle$ & \\
		& & $|n,- \rangle \rightarrow |n+W - 2,- \rangle$ & \\
  \hline 
	\end{tabular}
	\caption{ESR matrix elements for transitions between precessing Landau level eigenstates in a spin-orbit field which rotates about the $z$-axis with winding number $W$. The matrix element is given in terms of the angle $\zeta$ between spin polarization and the $z$-axis, and in units $ (\frac{g \mu_B |B_\parallel|}{2})^2$ for electrons and  $(2 e \alpha_H B_z |B_\parallel| n)^2$ for holes. In the case of a static in-plane magnetic field ($W = 2$), the ESR matrix element depends on the angle $\phi - \varphi$ between the static and oscillating magnetic fields.}
\end{table*}

In the past, both cyclotron resonance and electron spin resonance (ESR) have been used to study the band parameters of 2D semiconductor systems\cite{ESRCR1,ESRCR2,ESRCR3,ESRCR4}. In the absence of spin-orbit interaction, ESR occurs when the frequency of the applied in-plane magnetic field coincides with the energy splitting between Zeeman states belonging to the orbital level, and therefore simply measures the $g$-factor. In the presence of spin-orbit coupling, an oscillating in-plane magnetic field may result in transitions between different orbital levels, and we expect ESR to be observed in the same range of frequencies as cyclotron resonance.

The ESR probability depends on the angle $\zeta$ (\ref{zeta}) which describes mixing between $|+\rangle$ and $|-\rangle$ states in the precessing Landau level wavefunctions. Thus while magnetic oscillations offer a sensitive probe of the phase of the eigenvalues of the matrix phase $U(2\pi)$, ESR may provide a complementary measurement in the sense that it probes the spin structure of the Landau level eigenstates.

 Let us first consider the situation in electron systems. The probability of transition between different levels, $n \gg 1$ may be calculated from the semiclassical wavefunctions (\ref{rotwavef1}), (\ref{rotwavef2}). An oscillating magnetic field applied in the 2D plane, $\delta H \propto b_x \sigma_x + b_y \sigma_y$ generates transitions with probability amplitude
\begin{gather}
\langle n' s' | b_x \sigma_x + b_y \sigma_y | n s \rangle = \nonumber \\ \int{ \left[ \chi^\dagger_{s'}(\theta) (b_x \sigma_x + b_y \sigma_y) \chi_s(\theta)\right] e^{i (n - n')\theta} \frac{ d\theta}{2\pi}}
\end{gather}
where $\chi_s = \chi_{ns} \approx \chi_{n's}$, since $n,n'$ are large.
There are transitions within the same orbital level ($\psi_{n+} \rightarrow \psi_{n-}$), as well as transitions between different spin states in different orbital levels, ($\psi_{n+} \rightarrow \psi_{n+2W,-}$ for $-\frac{ W \omega}{2} - \frac{ g \mu_B B_z}{2} > 0$ and $\psi_{n+} \rightarrow \psi_{n-2W,-}$ for $-\frac{ W \omega}{2} - \frac{ g \mu_B B_z}{2} < 0$). In addition, there exist purely orbital transitions ($\psi_{n+} \rightarrow \psi_{n \pm W, +}$ and $\psi_{n-} \rightarrow \psi_{n\pm W, -}$). The transition probabilities are summarized in Table 1.

Let us now consider the case for hole systems. The transition matrix element is given by
\begin{gather}
- \alpha_H \langle n's'| b_+ \pi_+^2 \sigma_- +b_- \sigma_-^2 \sigma_+ |n s \rangle = \nonumber \\
- 2e \alpha_H B_z n \int{ \chi^\dagger_{s'}(\theta) \left[ b_+ e^{2i \theta} \sigma_- + b_- e^{-2 i \theta}\sigma_+ \right] \chi_s(\theta) e^{i (n'-n)\theta} \frac{d\theta}{2\pi}} \  \ .
\end{gather}
We obtain transitions within the same orbital level only in the case $W = 1$ (corresponding, e.g. to the (100) Dresselhaus interaction). There are transitions between opposite spin states ($\psi_{n+} \rightarrow \psi_{n - 2W + 2, - }, \psi_{n+} \rightarrow \psi_{n-2,-}$ for $\frac{ W \omega}{2}- \frac{ g \mu_B B_z}{2} > 0$ and $\psi_{n+} \rightarrow \psi_{n+2W - 2, -}, \psi_{n+} \rightarrow \psi_{n+2,-}$ for $\frac{ W \omega}{2} - \frac{g\mu_B B_z}{2} < 0$), as well as purely orbital transitions ($ \psi_{n+} \rightarrow \psi_{n+2 - W, +}, \psi_{n- } \rightarrow \psi_{n+W-2,-}$). These results are summarized in Table 1. For the case $W = 2$, the probability of transition between opposite spin states exhibits a dependence on the direction of the oscillating magnetic field. Let us consider the case when a static magnetic field $B_\parallel = (B_x, B_y)$ is present. The matrix element for the transition  is
\begin{gather}
\langle \psi_{n-2}, - | \delta H | \psi_{n,+} \rangle \propto 1 - \frac{ \sin^2 \zeta}{2} \left[ 1 + \cos 2 ( \phi - \varphi)\right]
\end{gather}
where $\phi - \varphi$ is the angle between the static and oscillating magnetic field. These transitions are indicated by the vertical arrows in Fig. 3.

\section{Conclusion}

We have obtained a semiclassical expression for the Landau level spectrum in a 2D spin-orbit coupled system via the one-dimensional Schr\"{o}dinger equation (\ref{spinevolution}) describing spin evolution around the cyclotron trajectory. In the Born-Oppenheimer approximation, the semiclassical quantization condition is strongly modified by spin dynamics  and the Landau level problem becomes equivalent to that of calculating the SU(2) matrix $U(\theta)$ associated with spin precession around a momentum space orbit of fixed radius $|\pi| = \sqrt{|2e B_z|(n+\frac{1}{2})}$. In the semiclassical quantization condition, the eigenvalues of $U(2\pi)$ constitutes the spin-dependent correction to the phase, and in the case of a rotating spin-orbit interaction, contains a geometric contribution which is associated with the out-of-plane tilting of the precessing spin wavefunctions relative to the driving spin-orbit field $\beta$. The importance of the geometric contribution in the experimental regime was illustrated for both $n$ and $p$ type systems in Figs. 2,5,7.
We have shown that magnetic oscillations directly probe the phase $\Phi$ of the eigenvalues $U(2\pi)$, while in the rotating case ESR measures the polarization of precessing Landau states. When spin dynamics is controlled by variation of external parameters such as the external magnetic field and gate voltage, this allows mapping of spin precession along the classical orbit.

\appendix

\section{Choice of Landau level labeling}
\cblack

In this appendix we present a  more detailed  derivation of the semiclassical solutions (\ref{rotwavef1}), (\ref{rotwavef2}) and demonstrate the choice for labeling Landau levels. \cblack 
The semiclassical formalism for the wavefunction $\psi(\theta)=e^{-i\eta n\theta}\chi(\theta)$  satisfies Eq. (12) for $\chi(\theta)$
\begin{gather}
-i\eta\omega\frac{\partial\chi}{\partial\theta}=[\beta(\theta,n)\cdot\sigma-\omega\delta]\chi
\label{eq12}
\end{gather}
where the examples in this work have the form $\beta(\theta,n)=[\beta_\parallel\cos(W\theta+\phi),\beta_\parallel\sin(W\theta+\phi),\beta_z]$, $W$ is an integer and $\beta_\parallel$ is implicitly $n$ dependent. Using the unitary transformation $g(\theta)=e^{-\frac{1}{2} iW\theta\sigma_z}$ and
$\chi(\theta)=e^{-\frac{1}{2} i\phi\sigma_z} g(\theta)\chi'(\theta)$  we simplify the interaction term and the equation for $\chi'(\theta)$ (similar to Eq. (\ref{corot}))
\begin{gather}
e^{\frac{i}{2} (W\theta+\phi)\sigma_z} (\beta(\theta)\cdot\sigma) e^{-\frac{i}{2} (W\theta+\phi)\sigma_z}=
\beta_\parallel\sigma_x+\beta_z\sigma_z\nonumber\\
\rightarrow  -i\eta\omega\frac{\partial\chi'(\theta)}{\partial\theta}=
\left[\beta_\parallel\sigma_x+(\beta_z+\frac{1}{2}\eta\omega W)\sigma_z -\omega\delta \right]\chi'(\theta)
\end{gather}  
 We next rotate around the y axis by the angle $\zeta$ (Eq. \ref{zeta}) where $\cos\zeta=\beta_\parallel/|\mathcal{B}|,\, \sin\zeta=(\beta_z+\frac{1}{2}\eta\omega W)/|{\mathcal B}|$ and $|{\mathcal B}|=\sqrt{\beta_\parallel^2+(\beta_z+\frac{1}{2}\eta\omega W)^2}$, \cblack
 so that the equation for $\chi''(\theta)=e^{\frac{1}{2}i \zeta\sigma_y}\chi'(\theta)$ becomes
\begin{gather}
-i\eta\omega\frac{\partial(\chi''(\theta))}{\partial\theta}=\left[\mathcal\sigma_z-\omega\delta \right]\chi''(\theta)\nonumber\\
\rightarrow \chi''(\theta) \sim e^{i\eta(\frac{1}{\omega}|\mathcal{B}|\sigma_z-\delta)\theta}
\end{gather}
This latter form multiplies any $\theta$ independent spinor, choosing the states $(1,0),\,(0,1)$ we find the solutions
\begin{gather}
\chi_+(\theta)=C_+
e^{i\eta(\frac{1}{\omega}|{\mathcal B}|-\delta)\theta}\left(   \begin{array}{c}e^{-\frac{1}{2}i (W\theta+\phi)}\cos\frac{\zeta}{2} \\
e^{\frac{1}{2}i (W\theta+\phi)}\sin\frac{\zeta}{2} \end{array}\right)\nonumber\\
\chi_-(\theta)=C_-
e^{i\eta(-\frac{1}{\omega}|{\mathcal B}|-\delta)\theta}\left(   \begin{array}{c}-e^{-\frac{1}{2}i (W\theta+\phi)}\sin\frac{\zeta}{2} \\
e^{\frac{1}{2}i (W\theta+\phi)}\cos\frac{\zeta}{2} \end{array}\right)\nonumber\\
\end{gather}
where the constants $C_\pm$ are $\theta$ independent.
Periodic boundary condition, i.e. uniqueness of wavefunction imply integers $m_\pm$, hence
two eigenvalues $\delta_\pm$, where
\begin{gather}
\eta(\pm |\mathcal{B}|-\omega\delta_\pm)\pm\frac{1}{2}  W\omega=m_\pm\omega\nonumber\\
\rightarrow \qquad \omega\delta_\pm=\pm(|\mathcal{B}|+\frac{1}{2}\eta W\omega)-m_\pm\omega
\end{gather}
and exponents with $\frac{1}{2} W\rightarrow-\frac{1}{2} W$ are also integers since the difference is an integer W. Hence
\begin{gather}
\chi_+(\theta)=C_+e^{im_+\theta}\left(   \begin{array}{c}e^{-i W\theta-\frac{i \phi}{2}}\cos\frac{1}{2}\zeta \\
e^{\frac{i\phi}{2}}\sin\frac{1}{2}\zeta \end{array}\right)\nonumber\\
\chi_-(\theta)=C_-e^{im_-\theta}\left(   \begin{array}{c}-e^{-\frac{i \phi}{2}}\sin\frac{1}{2}\zeta \\
e^{iW\theta +\frac{i \phi}{2}}\cos\frac{1}{2}\zeta \end{array}\right)
\end{gather}

It is interesting to note that while the full Hamiltonian (\ref{hamil}) is not time-reversal invariant, the one-dimensional equation for spin (\ref{eq12}) is invariant under the time-reversal operation $\mathcal{T} = i \sigma_y \mathcal{K}$ (where $\mathcal{K}$ is complex conjuation). Hence the solutions $\chi_\pm(\theta)$ are related by this operator so that $C_-=C_+^*$ and $m_-=-m_+$.
\cblack

The integers $m_\pm$ correspond to relabeling the Landau level index $n$ and within the semiclassical scheme any choice with $m_\pm\ll n$ is acceptable.  The energies are then given by
\begin{gather}
 E_n^\pm=\omega(n+\frac{1}{2})+\omega\delta_\pm \nonumber \\
=\omega(n-m_\pm+\frac{1}{2})\pm (|\mathcal{B}(n)|+\frac{1}{2}\eta W\omega)  \nonumber \\
 \rightarrow \omega(n' + \frac{1}{2}) \pm ( |\mathcal{B}(n' + m_\pm)| + \frac{1}{2} \eta W \omega)
 \end{gather}
where the relabelling $n \rightarrow n' = n + m_\pm$ only affects the lowest lying Landau levels which are not accessible in the semiclassical method.  In the semiclassical limit, $|\mathcal{B}(n')| \approx |\mathcal{B}(n)|, \ \ \ \zeta(n') \approx \zeta(n)$ for $m_\pm \ll n$, so the energies and wavefunctions are unchanged under the relabeling.  For the reasons stated in the text, we have chosen the solutions (\ref{rotwavef1}) corresponding to  $m_\pm = \pm W, \  C_\pm = e^{ \pm \frac{ i \phi}{2}}$, and (\ref{rotwavef2}) corresponding to $m_\pm = 0, \ C_\pm = e^{ \mp \frac{ i \phi}{2}}$, \cblack which minimise the leading error in the semiclassical scheme and therefore have best agreement with the exact solutions presented in Section IV.

\section{Exact spectra for $n$-type systems}


The Landau level spectrum of systems with a pure Rashba or Dresselhaus interaction may be solved by introducing creation and annihilation operators
\begin{gather}
\bm{a} = \frac{\bm{\pi}_-}{\sqrt{|2e B_z|}} \ \ , \ \ \bm{a}^\dagger = \frac{\bm{\pi}_+}{\sqrt{|2e B_z|}}
\end{gather}
and diagonalizing the Hamiltonian in the number basis $|n,\pm \rangle = |n\rangle |\pm\rangle$ where $\bm{a}^\dagger \bm{a}|n\rangle = n |n \rangle$, $\sigma_z |\pm \rangle = \pm |\pm \rangle$.

\subsection{Rashba interaction}
The Hamiltonian is
\begin{gather}
H = \omega( \bm{a}^\dagger \bm{a} + \frac{1}{2}) + \frac{i \alpha_R \sqrt{|2e B_z|}}{2}( \bm{a} \sigma_+ - \bm{a}^\dagger \sigma_-) - \frac{g\mu_B B_z}{2} \sigma_z \ \ . 
\end{gather}
The Rashba interaction $H_R \propto i\bm{a} \sigma_+ + h.c.$ couples basis states $|n,-\rangle$ and $|n-1,+\rangle$ for $n \geq 1$, with $|0,-\rangle$ being an eigenstate with energy $\frac{\omega}{2} + \frac{ g \mu_B B_z}{2}$. The remaining spectrum may be obtained by diagonalizing the $2\times 2$ Hamiltonian
\begin{gather}
H \rightarrow \left(\begin{array}{cc}
\omega(n + \frac{1}{2}) + \frac{g \mu_B B_z}{2} & -i \alpha_R \sqrt{|2e B_z|n} \\
i \alpha_R \sqrt{|2e B_z|n} & \omega(n - \frac{1}{2}) - \frac{g \mu_B B_z}{2} \end{array}\right)
\end{gather}
in the basis $(|n,-\rangle, |n-1,+\rangle)$ for $n \geq 1$, which gives energies
\begin{gather}
E_{n,\pm} = \omega n \pm \sqrt{( \frac{\omega}{2} + \frac{g \mu_B B_z}{2})^2 + |2e B_z|\alpha_R^2 n} \ \ .
\end{gather}
When $\frac{\omega}{2} + \frac{g \mu_B B_z}{2} > 0$, the energy of the eigenstate $|0,-\rangle$ coincides with $E_{0,+}$; in the opposite situation it coincides with $E_{0,-}$. Therefore the complete spectrum is given by
\begin{gather}
E_{n,+} \ \ ,  \ \ n = 0, 1, 2, \dots \nonumber \\
E_{n,-} \ \ , \ \ n = 1, 2, \dots 
\label{nrashbacond1}
\end{gather}
for $\frac{\omega}{2} + \frac{g \mu_B B_z}{2} > 0$, and
\begin{gather}
E_{n,+} \ \ ,  \ \ n = 1, 2, \dots \nonumber \\
E_{n,-} \ \ , \ \ n = 0, 1, 2, \dots 
\label{nrashbacond2}
\end{gather}
for $\frac{\omega}{2} + \frac{g \mu_B B_z}{2} < 0$. The eigenstates are given by
\begin{gather}
\psi_{n+} = \cos \frac{\zeta_n}{2} |n-1,+\rangle + i\sin \frac{ \zeta_n}{2} |n,-\rangle \ \ ,
\nonumber \\
\psi_{n-} = - \sin \frac{\zeta_n}{2} |n-1,+\rangle + i \cos \frac{\zeta_n}{2} |n,-\rangle
\end{gather}
where
\begin{gather}
\tan\zeta_n = \frac{ \alpha_R\sqrt{|2e B_z|n}}{ - \frac{\omega}{2} - \frac{g \mu_B B_z}{2}} \ \ ,
\end{gather}
and $n$ takes the same values as in the expressions (\ref{nrashbacond1}), (\ref{nrashbacond2}).

The wavefunctions $\psi_{n,\pm}(\theta) = \langle \theta| \psi_{n,\pm}\rangle$ in the $\theta$-representation may be obtained by use of Eq. (\ref{basis}); we obtain
\begin{gather}
\psi_{n,+}(\theta) = e^{i n\theta}( \cos \frac{ \zeta_n}{2} e^{-i \theta} |+\rangle +i \sin \frac{ \zeta_n}{2} | -\rangle) \ \ , \nonumber \\
\psi_{n,-}(\theta) = e^{in \theta} ( - \sin \frac{\zeta_n}{2} e^{-i \theta} |+\rangle + i\cos \frac{\zeta_n}{2} |-\rangle) \ \ .
\end{gather}
After a shift of index, $E_{n-} \rightarrow E_{n+1,-}$, $\psi_{n-} \rightarrow \psi_{n+1,-}$ for $\frac{ \omega}{2} + \frac{g \mu_B B_z}{2} > 0$ and $E_{n+} \rightarrow E_{n+1,+}$, $\psi_{n+}\rightarrow \psi_{n+1,+}$ for $\frac{ \omega}{2} + \frac{g \mu_B B_z}{2} < 0$ (so that the spectra in both spin series begin with index $n = 0$), we obtain the energies (\ref{spectrumRashba}) and wavefunctions (\ref{Rashbawave1}), (\ref{Rashbawave2}) shown in the text.

\subsection{Dresselhaus interaction}

The Hamiltonian is
\begin{gather}
H = \omega( \bm{a}^\dagger \bm{a} + \frac{1}{2}) + \frac{\alpha_D \sqrt{|2e B_z|}}{2} (\bm{a} \sigma_- + \bm{a}^\dagger \sigma_+) - \frac{g\mu_B B_z}{2} \sigma_z \ \ .
\end{gather}
the Dresselhaus interaction $H_D \propto \bm{a} \sigma_- + h.c.$ couples basis states $|n,+\rangle$ and $|n-1,-\rangle$ with $|0,+\rangle$ being an eigenstate with energy $\frac{\omega}{2} - \frac{g \mu_B B_z}{2}$. The remaining spectrum may be obtained by diagonalizing the $2\times 2$ Hamiltonian
\begin{gather}
H \rightarrow
\left(\begin{array}{cc}
\omega(n + \frac{1}{2}) - \frac{g \mu_B B_z}{2} & \alpha_D \sqrt{|2e B_z|n} \\
\alpha_D \sqrt{|2e B_z|n} & \omega(n - \frac{1}{2}) + \frac{g \mu_B B_z}{2}
\end{array}\right)
\end{gather}
in the basis $(|n,+\rangle, |n-1,-\rangle)$, which gives energies
\begin{gather}
E_{n,\pm} = \omega n \pm \sqrt{( \frac{\omega}{2} - \frac{g \mu_B B_z}{2})^2 + |2eB_z|\alpha_D^2 n} \ \ .
\end{gather}
When $\frac{\omega}{2} - \frac{g \mu_B B_z}{2} > 0$, the energy of the eigenstate $|0,+\rangle$ coincides with $E_{0,+}$; in the opposite situation it coincides with $E_{0,-}$. Therefore the complete spectrum is given by
\begin{gather}
E_{n,+} \ \ , \ \ n = 0, 1, 2, \dots \nonumber \\
E_{n,-} \ \ , \ \ n = 1, 2, \dots
\label{ndresscond1}
\end{gather}
for $ \frac{\omega}{2} - \frac{g \mu_B B_z}{2} > 0$, and
\begin{gather}
E_{n,+} \ \ , \ \ n = 1, 2, \dots \nonumber \\
E_{n,-} \ \ , \ \ n = 0, 1, 2, \dots
\label{ndresscond2}
\end{gather}
for $ \frac{\omega}{2} - \frac{g \mu_B B_z}{2} < 0$. The eigenstates are given by
\begin{gather}
\psi_{n+} = \cos \frac{\zeta_n}{2} |n,+\rangle + \sin \frac{ \zeta_n}{2} |n-1,-\rangle \ \ ,
\nonumber \\
\psi_{n-} = - \sin \frac{\zeta_n}{2} |n,+\rangle + \cos \frac{\zeta_n}{2} |n-1,-\rangle
\end{gather}
where
\begin{gather}
\tan\zeta_n = \frac{ \alpha_D\sqrt{|2e B_z|n}}{ \frac{\omega}{2} - \frac{g \mu_B B_z}{2}} \ \ ,
\end{gather}
and $n$ takes the same values as in the expressions (\ref{ndresscond1}), (\ref{ndresscond2}). The wavefunctions $\psi_{n,\pm}(\theta) = \langle \theta| \psi_{n,\pm}\rangle$ in the $\theta$-representation may be obtained by use of Eq. (\ref{basis}); we obtain
\begin{gather}
\psi_{n,+}(\theta) = e^{i n\theta}( \cos \frac{ \zeta_n}{2} |+\rangle + \sin \frac{ \zeta_n}{2} e^{-i \theta}| -\rangle) \ \ , \nonumber \\
\psi_{n,-}(\theta) = e^{in \theta} ( - \sin \frac{\zeta_n}{2}  |+\rangle + \cos \frac{\zeta_n}{2} e^{-i \theta} |-\rangle) \ \ .
\end{gather}
After a shift of index, $E_{n-} \rightarrow E_{n+1,-}$, $\psi_{n-} \rightarrow \psi_{n+1,-}$ for $\frac{ \omega}{2} - \frac{g \mu_B B_z}{2} > 0$ and $E_{n+} \rightarrow E_{n+1,+}$, $\psi_{n+}\rightarrow \psi_{n+1,+}$ for $\frac{ \omega}{2} - \frac{g \mu_B B_z}{2} < 0$ (so that the spectra in both spin series begin with index $n = 0$), we obtain the energies (\ref{dressexact}) and wavefunctions (\ref{wavedress1}), (\ref{wavedress2}) shown in the text.

\section{Exact spectra for $p$-type systems}

In the text, three situations are discussed: the case of an in-plane magnetic field, a pure Rashba interaction, and a pure Dresselhaus interaction. In the hole case the creation and annihilation operators are
\begin{gather}
\bm{a} = \frac{\bm{\pi}_+}{\sqrt{2e B_z}} \ \ , \ \ \bm{a}^\dagger = \frac{\bm{\pi}_+}{\sqrt{2e B_z}}
\end{gather}
(note that they are reversed in comparison to the electron case due to the opposite sign of the electric charge). As in the electron case we obtain analytical solutions by diagonalizing the Hamiltonian in the number representation.

\subsection{In-plane magnetic field}

The Hamiltonian is
\begin{gather}
H = \omega( \bm{a}^\dagger \bm{a} + \frac{1}{2}) + e B_z \alpha_H B_\parallel (e^{i \phi} \bm{a}^2 \sigma_- + e^{-i \phi} (\bm{a}^\dagger)^2 \sigma_+) \nonumber \\
- \frac{g\mu_B B_z}{2} \sigma_z \ \ .
\end{gather}
where the phase $\phi$ is related to the direction of the in-plane magnetic field via $B_+ = B_\parallel e^{i \phi}$. The in plane field, $H_Z \propto e^{i \phi}\bm{a}^2 \sigma_- + h.c.$ couples basis states $|n,+\rangle$ and $|n-2,-\rangle$ with $|0,+\rangle$ and $|1,+\rangle$ being eigenstates with energies $\frac{\omega}{2} - \frac{g \mu_B B_z}{2}$ and $\frac{3 \omega}{2} - \frac{g \mu_B B_z}{2}$ respectively. The remaining spectrum may be obtained by diagonalizing the $2\times 2$ Hamiltonian
\begin{gather}
H \rightarrow
\left(\begin{array}{cc}
\omega(n + \frac{1}{2}) - \frac{g \mu_B B_z}{2} & 2e \alpha_H B_z B_\parallel e^{-i \phi} \nu_n \\
2e \alpha_H B_z B_\parallel e^{ i \phi} \nu_n & \omega(n - \frac{3}{2}) + \frac{g \mu_B B_z}{2}
\end{array}\right) \ \ ,
\end{gather}
where
\begin{gather}
\nu_n = \sqrt{n(n-1)}
\end{gather}
in the basis $(|n,+\rangle, |n-2,-\rangle)$, which gives energies
\begin{gather}
E_{n,\pm} = \omega (n -\frac{1}{2}) \pm \sqrt{( \omega- \frac{g \mu_B B_z}{2})^2 + (2 e \alpha_H B_z B_\parallel \nu_n)^2} \ \ .
\end{gather}
When $\omega - \frac{g \mu_B B_z}{2} > 0$, the energy of the eigenstates $|0,+\rangle$ and $|1,+\rangle$ coincide with $E_{0,+}$ and $E_{1,+}$ respectively; in the opposite situation they coincide  with $E_{0,-}, E_{1,-}$. Therefore the complete spectrum is given by
\begin{gather}
E_{n,+} \ \ , \ \ n = 0, 1, 2, \dots \nonumber \\
E_{n,-} \ \ , \ \ n = 2,3, \dots
\label{inplanecond1}
\end{gather}
for $\omega - \frac{g \mu_B B_z}{2} > 0$, and
\begin{gather}
E_{n,+} \ \ , \ \ n = 2, 3,\dots \nonumber \\
E_{n,-} \ \ , \ \ n = 0, 1, 2, \dots
\label{inplanecond2}
\end{gather}
for $\omega - \frac{g \mu_B B_z}{2} < 0$. The eigenstates are given by
\begin{gather}
\psi_{n+} = \cos \frac{\zeta_{\nu_n}}{2} |n,+\rangle + \sin \frac{ \zeta_{\nu_n}}{2} e^{i \phi}|n-2,-\rangle \ \ ,
\nonumber \\
\psi_{n-} = - \sin \frac{\zeta_{\nu_n}}{2} |n,+\rangle + \cos \frac{\zeta_{\nu_n}}{2}e^{i \phi} |n-2,-\rangle
\end{gather}
where
\begin{gather}
\tan\zeta_{\nu_n} = \frac{ 2e \alpha_H B_z B_\parallel \nu_n}{ \omega - \frac{g \mu_B B_z}{2}} \ \ ,
\end{gather}
and $n$ takes the same values as in the expressions (\ref{inplanecond1}), (\ref{inplanecond2}). The wavefunctions $\psi_{n,\pm}(\theta) = \langle \theta| \psi_{n,\pm}\rangle$ in the $\theta$-representation may be obtained by use of Eq. (\ref{basis}); we obtain
\begin{gather}
\psi_{n,+}(\theta) = e^{-i n\theta}( \cos \frac{ \zeta_{\nu_n}}{2} |+\rangle + \sin \frac{ \zeta_n}{2} e^{2i \theta}| -\rangle) \ \ , \nonumber \\
\psi_{n,-}(\theta) = e^{-in \theta} ( - \sin \frac{\zeta_n}{2}  |+\rangle + \cos \frac{\zeta_n}{2} e^{2i \theta} |-\rangle) \ \ .
\end{gather}
After a shift of index, $E_{n-} \rightarrow E_{n+2,-}$, $\psi_{n-} \rightarrow \psi_{n+2,-}$ for $\omega - \frac{g \mu_B B_z}{2} > 0$ and $E_{n+} \rightarrow E_{n+2,+}$, $\psi_{n+}\rightarrow \psi_{n+2,+}$ for $\omega - \frac{g \mu_B B_z}{2} < 0$ (so that the spectra in both spin series begin with index $n = 0$), we obtain the spectrum (\ref{inplanepexact}) and wavefunctions (\ref{inplanewave1}), (\ref{inplanewave2}) shown in the text.

\subsection{Rashba interaction}

The Hamiltonian is
\begin{gather}
H = \omega( \bm{a}^\dagger \bm{a} + \frac{1}{2}) + \frac{ i \alpha'_R (2e B_z)^\frac{3}{2}}{2} ( \bm{a}^3 \sigma_- - (\bm{a}^\dagger)^3 \sigma_+) \nonumber \\
- \frac{g \mu_B B_z}{2} \sigma_z \ \ .
\end{gather}
The Rashba interaction $H'_R \propto \bm{a}^3 \sigma_-$ couples basis states $|n,+\rangle$ and $|n-3,-\rangle$ for $n \geq 3$. For $n = 0, 1, 2$, the basis states $|n,+\rangle$ are eigenstates with energy $\omega(n + \frac{1}{2}) - \frac{g \mu_B B_z}{2}$. The remaining spectrum may be obtained by diagonalizing the $2\times 2$ Hamiltonian
\begin{gather}
H \rightarrow\left(\begin{array}{cc}
\omega(n + \frac{1}{2}) - \frac{ g\mu_B B_z}{2} & -i \alpha'_R (2e B_z \nu_n)^3 \\
i \alpha'_R (2e B_z \nu_n)^3 & \omega(n - \frac{5}{2}) + \frac{g \mu_B B_z}{2}) \end{array}
\right)
\end{gather}
where
\begin{gather}
\nu_n = ( n (n-1)(n-2))^\frac{1}{3}
\end{gather}
in the basis $(|n,+\rangle, |n-3,-\rangle)$, giving energies
\begin{gather}
E_{n,\pm} = \omega( n - 2) \pm \sqrt{( \frac{3 \omega}{2} - \frac{g \mu_B B_z}{2})^2 + (\alpha'_R)^2 (2e B_z \nu_n)^3} \ \ .
\end{gather}
When $ \frac{3 \omega}{2} - \frac{g \mu_B B_z}{2} > 0$, the energies of the eigenstates $|n,+\rangle$ for $n = 0,1, 2$ coincide with $E_{n,+}$; in the opposite situation they coincide with $E_{n,-}$. Therefore the complete spectrum is given by
\begin{gather}
E_{n,+} \ \ , \ \ n = 0, 1, 2, \dots \nonumber \\
E_{n,-} \ \ , \ \ n = 3, 4, 5, \dots
\label{prashbacond1}
\end{gather}
for $ \frac{3 \omega}{2} - \frac{g \mu_B B_z}{2} > 0$ and
\begin{gather}
E_{n,+} \ \ , \ \ n = 3, 4, 5, \dots \nonumber \\
E_{n,-} \ \ , \ \ n = 0, 1, 2, \dots 
\label{prashbacond2}
\end{gather}
for $\frac{3 \omega}{2} - \frac{g \mu_B B_z}{2} < 0$. The eigenstates are given by
\begin{gather}
\psi_{n,+} = \cos \frac{\zeta_{\nu_n}}{2} |n,+ \rangle + i \sin \frac{ \zeta_{\nu_n}}{2} |n-3,-\rangle \nonumber \\
\psi_{n,-} = - \sin \frac{\zeta_{\nu_n}}{2} |n,+\rangle + i \cos \frac{ \zeta_{\nu_n}}{2} |n-3,-\rangle
\label{rashbapstate}
\end{gather}
where
\begin{gather}
\tan \nu_n = \frac{ \alpha'_R (2 e  B_z \nu_n)^\frac{3}{2}}{ \frac{3 \omega}{2} - \frac{g \mu_B B_z}{2}} \ \ ,
\end{gather}
and $n$ takes the same values as in the expressions (\ref{prashbacond1}), (\ref{prashbacond2}).

The wavefunctions the $\theta$-representation are given by projection of the states (\ref{rashbapstate}) onto the basis (\ref{basis}),
\begin{gather}
\psi_{n,+}(\theta) = e^{-i n \theta} ( \cos \frac{ \zeta_{\nu_n}}{2} |+\rangle + i \sin \frac{\zeta_{\nu_n}}{2} e^{3i \theta} |-\rangle) \ \ ,
\nonumber \\
\psi_{n, -}(\theta) = e^{-i n \theta} ( - \sin \frac{ \zeta_{\nu_n}}{2} |+\rangle + i \cos \frac{\zeta_{\nu_n}}{2} e^{3i \theta} |-\rangle) \ \ .
\end{gather}
 After a shift of index, $E_{n-} \rightarrow E_{n+2,-}$, $\psi_{n-} \rightarrow \psi_{n+3,-}$ for $\frac{3 \omega}{2} - \frac{g \mu_B B_z}{2} > 0$ and $E_{n,+} \rightarrow E_{n+2,+}$, $\psi_{n,+}\rightarrow \psi_{n+3,+}$ for $\frac{3 \omega}{2} - \frac{g \mu_B B_z}{2} < 0$, we obtain the energies (\ref{rashbapexact}) and wavefunctions (\ref{rashbapwave1}), (\ref{rashbapwave2}) shown in the text.

\subsection{Dresselhaus interaction}

The Hamiltonian is
\begin{gather}
H = \omega( \bm{a}^\dagger \bm{a} + \frac{1}{2}) 
- \frac{g \mu_B B_z}{2} \sigma_z + \nonumber \\
\frac{\alpha'_D(2e B_z)^\frac{3}{2}}{4} ( (\bm{a}^2 \bm{a}^\dagger + \bm{a}^\dagger \bm{a}^2 ) \sigma_- + ((\bm{a}^\dagger)^2 \bm{a} + \bm{a} (\bm{a}^\dagger)^2) \sigma_-) 
\end{gather}
where $\alpha'_D$ is the Dresselhaus constant for the heterostructure. The Dresselhaus interaction couples basis states $|n,+\rangle$ and $|n-1,-\rangle$ for $n \geq 1$. The state $|0,+\rangle$ is an eigenstate with energy $\frac{\omega}{2} - \frac{ g \mu_B B_z}{2}$. The remaining spectrum is given by diagonalization of the $2\times 2$ Hamiltonian
\begin{gather}
H \rightarrow \left(\begin{array}{cc}\omega(n + \frac{1}{2}) - \frac{g \mu_B B_z}{2} & \alpha'_D (2e B_z n)^\frac{3}{2} \\
\alpha'_D(2e B_z n)^\frac{3}{2} & \omega(n - \frac{1}{2}) + \frac{g \mu_B B_z}{2}
\end{array}\right) \ \ .
\end{gather}
in the basis $(|n,+\rangle, |n-1,-\rangle)$. The energies are given by
\begin{gather}
E_{n+} = \omega n \pm \sqrt{( \frac{\omega}{2} - \frac{g \mu_B B_z}{2})^2 + (\alpha'_D)^2 (2e B_z n)^3} \ \ .
\end{gather}
When $\frac{\omega}{2} - \frac{g \mu_B B_z}{2} > 0$, the energy of the state $|0,+\rangle$ coincides with $E_{0,+}$. In the opposite situation, it coincides with $E_{0,-}$. Thus the complete energy spectrum consists of
\begin{gather}
E_{n,+} \ \ , \ \ n = 0, 1, 2, \dots \nonumber \\
E_{n,-} \ \ ,  \ \ n = 1, 2, \dots
\label{dresspcond1}
\end{gather}
for $\frac{\omega}{2} - \frac{g \mu_B B_z}{2} > 0$, and
\begin{gather}
E_{n,+} \ \ , \ \ n = 1, 2, \dots \nonumber \\
E_{n,-} \ \ , \ \ n= 0, 1, 2, \dots
\label{dresspcond2}
\end{gather}
for $\frac{\omega}{2} - \frac{g \mu_B B_z}{2} < 0$. The eigenstates are given by
\begin{gather}
\psi_{n,+} = \cos \frac{\zeta_n}{2} |n, + \rangle + \sin \frac{\zeta_n}{2} |n-1,-\rangle \nonumber \\
\psi_{n,-} = - \sin \frac{ \zeta_n}{2} | n, +\rangle + \cos \frac{\zeta_n}{2} |n - 1, -\rangle \ \ 
\label{dresspstates}
\end{gather}
where
\begin{gather}
\tan \zeta_n =  \frac{ \alpha'_D (2e B_z n)^\frac{3}{2}}{ \frac{\omega}{2} - \frac{g \mu_B B_z}{2}} \ \ ,
\end{gather}
and $n$ takes the same values as in the expressions (\ref{dresspcond1}), (\ref{dresspcond2}). The wavefunctions in the $\theta$-representation are given by projection of the states (\ref{dresspstates}) onto the basis (\ref{basis}),
\begin{gather}
\psi_{n,+}(\theta) = e^{-i n \theta} (\cos \frac{ \zeta_n}{2} |+\rangle  + \sin \frac{\zeta_n}{\theta} e^{i \theta} |-\rangle) \ \ , \nonumber \\
\psi_{n,-} (\theta) = e^{-i n \theta} ( - \sin \frac{ \zeta_n}{2} |+\rangle + \cos \frac{ \zeta_n}{2} e^{i \theta} |-\rangle) \ \ .
\end{gather}
  After a shift of index, $E_{n-} \rightarrow E_{n+1,-}$, $\psi_{n,-}\rightarrow \psi_{n+1,-}$ for $\frac{ \omega}{2} - \frac{g \mu_B B_z}{2} > 0$ and $E_{n,+} \rightarrow E_{n+1,+}$, $\psi_{n,+} \rightarrow \psi_{n+1,+}$ for $\frac{\omega}{2} - \frac{g \mu_B B_z}{2} < 0$, we obtain the energies (\ref{rashbapexact}) and wavefunctions (\ref{rashbapwave1}), (\ref{rashbapwave2}) shown in the text.

\end{document}